\documentclass[utf8]{frontiersSCNS} 
\usepackage{url,hyperref,lineno,microtype,subcaption}
\usepackage[onehalfspacing]{setspace}

\DeclareMathOperator*{\POPCOUNT}{POPCOUNT}

\newcommand \changed {}
\newcommand \newchanged {}

\def\keyFont{\fontsize{8}{11}\helveticabold }
\def\firstAuthorLast{Hirtzlin {et~al.}} 
\def\Authors{
 Tifenn~Hirtzlin\,$^{1,\dagger}$, 
 Marc~Bocquet\,$^{2,\dagger}$,
 Bogdan~Penkovsky\,$^{1}$,
 Jacques-Olivier~Klein\,$^{1}$, 
 Etienne~Nowak\,$^{3}$, 
 Elisa~Vianello\,$^{3}$, 
 Jean-Michel~Portal\,$^{2}$ 
 and Damien~Querlioz\,$^{1,*}$}
\def\Address{$^{1}$ C2N, Univ Paris-Sud, Universit\'e Paris-Saclay, CNRS, Palaiseau, France \\
$^{2}$ Aix Marseille Univ, Universit\'e de Toulon, CNRS, IM2NP, Marseille, France \\ 
$^{3}$ CEA, LETI, Grenoble, France  \\
$^{\dagger}$ Tifenn Hirtzlin and Marc Bocquet contributed equally to this work.}        

\def\corrAuthor{Damien Querlioz}
\def\corrEmail{damien.querlioz@c2n.upsaclay.fr}

\begin{document}
\onecolumn
\firstpage{1}
\title[Digital Biologically Plausible Implementation...]
{Digital Biologically Plausible Implementation of Binarized Neural Networks with Differential Hafnium Oxide Resistive Memory Arrays} 

\author[\firstAuthorLast ]{\Authors} 
\address{} 
\correspondance{} 

\extraAuth{}

\maketitle

\begin{abstract}

\section{}
\newchanged{The brain performs} intelligent tasks with extremely low energy consumption. 
\changed{
This work takes inspiration from two strategies used \newchanged{by the brain} to achieve this energy efficiency:  the absence of separation between computing and memory functions, and the reliance on low precision computation.}
The emergence of resistive memory technologies indeed provides an opportunity to \newchanged{co-integrate tightly} logic and memory  in hardware. In parallel, the recently proposed concept of Binarized Neural Network, where multiplications are replaced by exclusive NOR (XNOR) logic gates, \newchanged{offers} a way to implement artificial intelligence using very low precision computation. In this work, we therefore propose a strategy to implement low energy Binarized Neural Networks, which employs \changed{brain-inspired} \newchanged{concepts}, while retaining energy benefits from digital electronics. We design, fabricate and test a memory array, including periphery and sensing circuits, optimized for this in-memory computing scheme. Our circuit employs hafnium oxide resistive memory integrated in the back end of line of a 130 nanometer CMOS process, in a two transistors - two resistors cell, which allows performing the exclusive NOR operations of the neural network directly within the sense amplifiers.  We show, based on extensive electrical measurements, that our design allows reducing the amount of bit errors on the synaptic weights, without the use of formal error correcting codes. We design a whole system using this memory array. We show on standard machine learning tasks (MNIST, CIFAR-10, ImageNet and an ECG  task) that the system has an inherent resilience to bit errors. We evidence that its energy consumption is attractive \newchanged{compared to} more standard approaches, and  that it can use the memory devices in regimes where they exhibit particularly low programming energy and high endurance. 
We conclude the work by discussing   \newchanged{how it associates  biologically-plausible ideas with}  more traditional digital electronics concepts.

\tiny
 \keyFont{ \section{Keywords:} binarized neural networks, resistive memory, memristor, in-memory computing, biologically plausible digital electronics, ASICs.} 
\end{abstract}

\section{Introduction}

Through the progress of deep learning, artificial intelligence has made tremendous achievements in recent years.
Its energy consumption on graphics or central processing units (GPUs and CPUs) remains, however, a considerable challenge, limiting its use at the edge and raising the question of the sustainability of large scale artificial intelligence-based services.
Brains, by contrast, manage intelligent tasks with highly reduced energy usage.
One key difference between GPUs and CPUs on \newchanged{one hand}, and brains \newchanged{on the other hand} is \newchanged{how they deal} with memory.
In GPUs and CPUs, memory and arithmetic units are separated, both physically and conceptually.
In artificial intelligence algorithms, which require \newchanged{large} amounts of memory access, considerably more energy is spent moving data between logic and memory than for doing actual arithmetic \citep{pedram2017dark}.
In brains, by contrast, neurons -- which implement most of the arithmetic -- and synapses -- which are believed to store long term memory -- are entirely colocated.
A major lead for reducing the energy consumption of artificial intelligence is therefore to imitate this strategy and design non-von Neumann systems where logic and memory are merged \citep{yu2018neuro,editorial_big_2018,querlioz2015bioinspired,indiveri2015memory}. 
This idea 
\newchanged{
experiences a renewed interest today with the advent of
}
novel nanotechnology-based non-volatile memories,
which are compact and fast,
and can be embedded at the core of the Complementary Metal Oxide Semiconductor (CMOS) process  \citep{yu2018neuro,ambrogio2018equivalent,prezioso2015training,serb2016unsupervised, saighi2015plasticity,covi2016analog,wang20152}.
Another key difference between processors and the brain is the basic nature of computations. 
GPUs and CPUs typically perform all neural networks computations with precise floating point arithmetic.
In brains, most of the computation is done in a low precision analog fashion within the neurons \changed{\citep{faisal2008noise,klemm2005topology}}, resulting in asynchronous spikes 
\newchanged{as an output, which is therefore binary}. 
A second idea for \newchanged{cutting} the energy consumption of artificial intelligence therefore is to design systems that \newchanged{operate} with much lower precision computation.

In recent years, considerable research has proposed to implement neural networks using analog resistive memory as synapses -- the device conductance implementing the synaptic weights. 
\newchanged{To a large extent, neural network computation can be done} using analog electronics: weight / neuron multiplication is performed \newchanged{out of} Ohm's law, and addition is 
\newchanged{natively implemented with}
Kirchoff's current law \citep{ambrogio2018equivalent,prezioso2015training,serb2016unsupervised,li2018efficient,wang2018fully,shafiee2016isaac}.
 This type of implementation is to a certain extent very biologically plausible, as it reproduces the two strategies mentioned above.
 \newchanged{The very challenge of this implementation, however,} is that \newchanged{it requires} to be associated with relatively heavy analog or mixed-signal CMOS circuitry such as operational amplifier or Analog to Digital Converters, 
 \newchanged{resulting in significant}
 area and energy overhead.

In parallel, a novel class of neural networks has recently been proposed -- Binarized Neural Networks (or the closely related XNOR-NETs)  \citep{courbariaux2016binarized,rastegari2016xnor}.
In these neural networks, once trained, synapses as well as neurons assume only  binary values meaning $+1$ or $-1$.
These neural networks have therefore limited memory requirements, and also rely on highly simplified \newchanged{arithmetics}. 
In particular, multiplications are replaced by one-bit exclusive NOR (XNOR) operations. 
Nevertheless, \changed{Binarized Neural Networks} can achieve near state of the art performance on vision tasks \citep{courbariaux2016binarized,rastegari2016xnor,lin2017towards}, 
 and are therefore extremely attractive for realizing inference hardware.
The low precision of Binarized Neural Networks, and in particular the binary nature of neurons -- which is reminiscent of biological neurons spikes -- also 
\newchanged{endows them with}
biological plausibility: \changed{they can indeed} be seen as a simplification of spiking neural networks.

\changed{Great efforts have been devoted} to develop hardware \newchanged{implementations} of Binarized Neural Networks. 
Using nanodevices, 
\newchanged{one natural intuition would be to}
adopt the same strategy proposed for conventional neural networks and perform arithmetic in an analog fashion using Kirchoff's law \citep{yu2018neuro,yu2016binary}.
However, Binarized Neural Networks are very digital in nature, and are multiplication-less. 
\changed{These networks can therefore provide} an opportunity \changed{to} benefit \changed{at the same time from bioinspired ideas}  and from the achievements of Moore's law and digital electronics. 
In this work we propose a fully digital implementation of binarized neural networks \changed{incorporating} CMOS and nanodevices, \changed{and implementing} the biological concepts of tight memory and logic integration, as well as low precision computing.
As memory nanodevices, we use hafnium oxide-based resistive random access memory (OxRAM), \newchanged{a compact and fast non-volatile memory cell} \changed{fully} compatible with the CMOS process \citep{grossi2016fundamental}.

\newchanged{However, one significant} challenge to implement a digital system with memory nanodevices is their inherent variability \citep{ielmini2018memory,ly2018role}, which causes bit errors.
Traditional \changed{memory applications} employ multiple error  correcting codes (ECCs) to solve this issue.
\changed{
ECC decoding circuits have \newchanged{large area and high energy consumption} \citep{gregori2003chip}, and add extra time to data access due to syndrome computation and comparison. Moreover, the arithmetic operations of error syndrome computation are actually more complicated \changed{than} those of a Binarized Neural Network.
This solution is difficult to implement in a context where memory and logic are tightly integrated specially when part of the computation is performed during sensing. It is one of the main  reason for which the state of the art of RRAM for in memory computing does not correct errors and is not compatible with technologies with errors
\citep{chen_65nm_2018,chen_16mb_2017}.  
}
In this paper, we introduce our solution. We design, fabricate, and test a differential oxide-based resistive memory array, including all peripheral and sensing circuitry. This array, based on two-transistors / two-resistors (2T2R) bit cell inherently reduces bit errors without the use of ECC, and we show that it is particularly adapted for in-memory computing.
\newchanged{Then, we} design and simulate a \changed{fully} binarized neural network based on this memory array. We show that the XNOR operations can be integrated directly within the sense operation of the memory array, and that the resulting system can be highly energy efficient.
Based on neural networks on multiple datatsets (MNIST, CIFAR-10, ImageNet and ECG data analysis), we evaluate the amount of bit errors in the memory that can be tolerated by the system. Based on this information, we show that the memory nanodevices can be used in an unconventional programming regime, where they feature low programming energy (less than five picoJoules per bit) and outstanding endurance (billions of cycles).

Partial and preliminary results of this work have been presented at a conference \citep{bocquet2018}. This paper adds additional  measurements of OxRAMs with shorter programming pulses, an analysis of the impact of bit errors on more  datasets (ImageNet and ECG data analysis), and a detailed comparison and benckmarking of our approach with processors, non binarized ASIC neural networks and analog RRAM-based neural networks.


\section{Materials and Methods}
\subsection{Differential Memory Array for In-Memory Computing}

\begin{figure}[h!]
\begin{center}
\includegraphics[width=16cm]{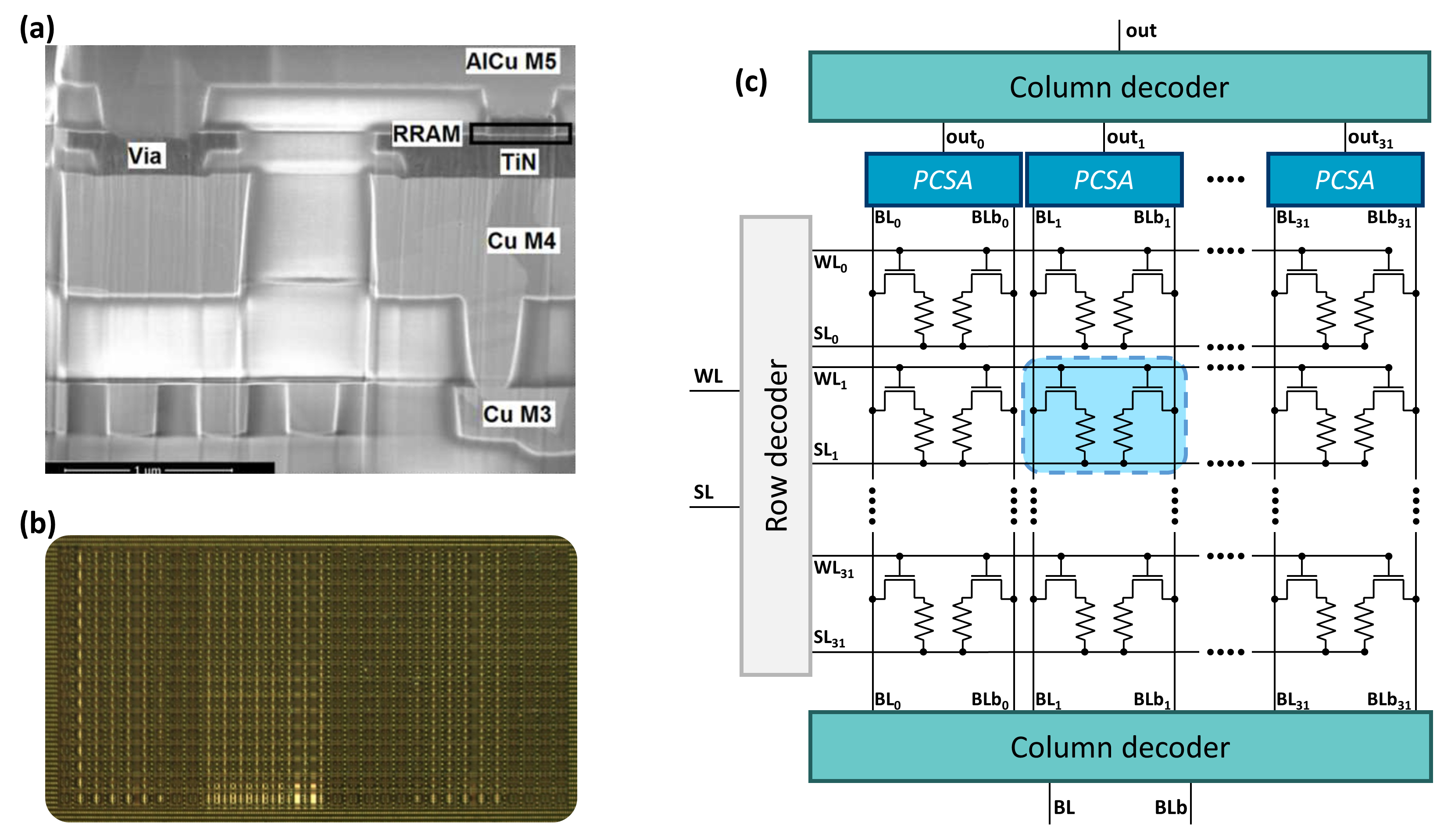} 
\end{center}
\caption{ (a) Scanning Electron Microscopy image of the \newchanged{back-end-of-line} of the CMOS process integrating an OxRAM device. (b) Photograph and (c) simplified schematic of the one kilobit  in-memory computing-targeted memory array characterized in this work.
}\label{fig:device_array}
\end{figure}

In this work, we fabricated a memory array for in-memory computing with its associated \newchanged{peripheral} and sensing circuits.
The memory cell relies on hafnium oxide ($\mathrm{HfO_2}$) oxide-based resistive Random Access Memory (OxRAM).
The  stack of the device is composed of a  $\mathrm{HfO_2}$ \newchanged{layer and a titanium layer. Both layers have a thickness of ten nanometers, and they grow between two} titanium nitride (TiN) electrodes. 
Our devices are embedded within the back-end-of-line of a commercial $130$ nanometer CMOS logic process (Fig.~\ref{fig:device_array}(a)), allowing tight integration of logic and non volatile memory \citep{grossi2016fundamental}. The devices are integrated  on top of the fourth (copper) metallic layer.

We chose hafnium oxide OxRAMs as they are known to provide non-volatile memories  compatible with modern CMOS process, and only involve foundry-friendly materials and process steps. 
After a one-time forming process, such devices can switch between low resistance and high resistance states (LRS and HRS) by applying positive or negative electrical pulses respectively.
\changed{
Our work could be reproduced with other types of  memories.
NOR flash cells, which are readily available in commercial process could  be used, and their potential for neuromorphic inference has been proven \cite{merrikh2017high}. However, they suffer from high programming voltages (higher than ten volts) requiring charge pumps, limited endurance, and \newchanged{they are not scalable} to the most advanced technology nodes \citep{dong201711}.
Emerging memories such as phase change memory or spin torque magnetoresitive memory could also be used using the strategies presented in this paper. These technologies do not require a forming process and \newchanged{they} can bring enhanced reliability with regards to OxRAMs, but come with an increased process cost  \citep{chen2016review}.
 }

Conventionally, OxRAMs are organized in a ``One Transistor - One Resistor'' structure (1T1R), where each nanodevice is associated with one access transistor \citep{chen2016review}.
\changed{The LRS and HRS are used to mean the zero and one logic values, or inversely.}
The read operation is then achieved by comparing the electrical resistance of the nanodevice to a reference value intermediate between typical values of resistances in HRS and LRS. 
Unfortunately, due to device variability, OxRAMs are prone to bit errors: 
the HRS value can become lower than the reference resistance, and the LRS value can be higher than the reference resistance. 
The device variability includes both device-to-device mismatch, as well as the fact that within the same device, the precise value of HRS and LRS resistance  changes at each programming cycle \citep{grossi2018experimental}. 

To limit the amount of bit errors, in this work, we fabricated a memory array with a ``Two Transistors - Two Resistors'' structure (2T2R), where each bit of information is stored in a pair of 1T1R structures.
A photograph of the die is presented in Fig.~\ref{fig:device_array}(b) and its simplified schematic in Fig.~\ref{fig:device_array}(c).
Information \changed{is  stored} in a differential fashion: the pair LRS/HRS means logic value zero, while the pair HRS/LRS means logic value one. In this situation, readout is performed by comparing the resistance values of the two devices. We therefore expect bit errors to be less frequent, as \changed{a bit error only occurs if} a device programmed in LRS \changed{is} more resistive than its complementary device programmed in HRS. 
This concept of 2T2R memory arrays has already been proposed,  but its benefit in terms of bit error rate has never been demonstrated until this work
\citep{hsieh_differential_2017,shih_twin-bit_2017}.

\begin{figure}[h!]
\begin{center}
\includegraphics[width=16cm]{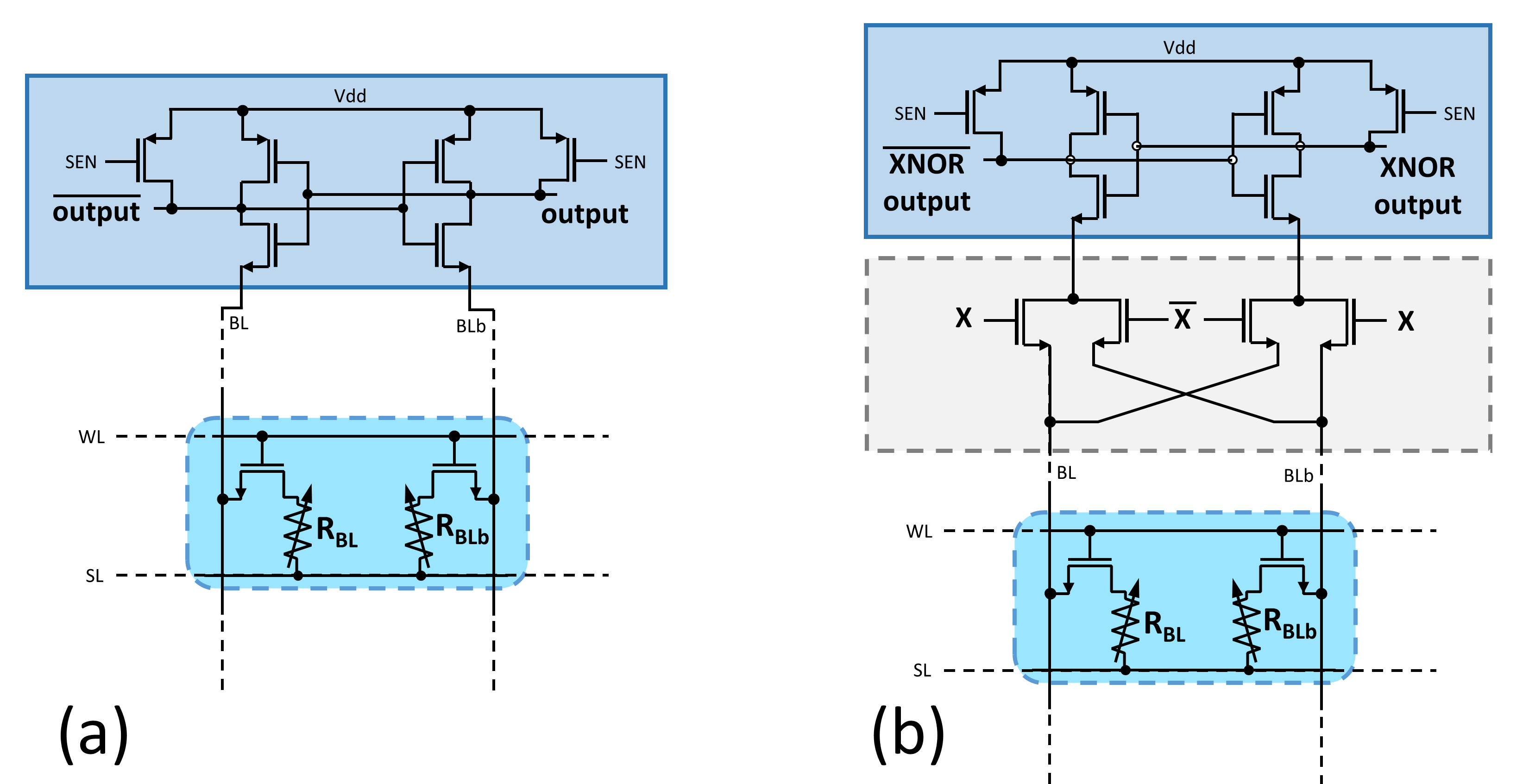} 
\end{center}
\caption{
(a) Schematic of the precharge sense amplifier used in this work to read 2T2R memory cells. (b) Schematic of the precharge sense amplifier augmented with a XNOR logic operation.}\label{fig:PCSAs}
\end{figure}

\changed{
The programming of devices in our array is made sequentially, i.e. on a device-by-device basis.
The first time that the memory array is used, all devices are ``formed''. To form the device of row $i$ and column $j$,  the  bit line $BL_j$, connected to the bottom electrode of the memory device, is set to ground, 
and the word line $WL_i$ is set to a voltage  chosen to limit the current to a ``compliance value'' of $200\mu A$.
A voltage ramp is applied to the sense line $SL_i$ connected to the top electrode of the memory device, increasing from $0$ to $3.3V$ at a ramp rate of $1000V/s$.
This forming operation is performed only once over the lifetime of the device.
To program a device into its LRS (SET operation),
the bit line $BL_j$ is set to ground, while the sense line $SL_i$ is set to $2V$. The word line $WL_i$ is again set to a voltage chosen to limit the current to a compliance value, ranging from $20\mu A$ to $200\mu A$, depending on the chosen programming condition. 
To program a device into its HRS (RESET operation), a voltage from opposite sign needs to be applied to the device, and the current compliance is not needed.    The sense line $SL_i$ is therefore set to the ground, while the word line $WL_i$ is set to a value of $3.3V$,  and the bit line $BL_j$ to a ``RESET voltage'' ranging from $1.5V$ to $2.5V$, depending on the chosen programming condition. 
For both SET and RESET operations, programming duration can range from $0.1\mu s$ to $100\mu s$.
During programming operations, all bit, select and word lines corresponding to non-selected devices are grounded, to the exception of the bit line of the complementary device of the selected device: this one is programmed to the same voltage as the one applied to the sense line, to avoid any disturb effect on the complementary device.
}

In our fabricated circuit, \changed{the} readout operation is performed with precharge sense amplifiers (PCSA) \citep{zhao2009high,zhao2014synchronous} (Fig.~\ref{fig:PCSAs}(a)). 
These circuits are highly energy efficient due to their operation in two phases, \changed{ precharge and discharge}, avoiding any direct path between supply voltage and ground.
First, the sense signal (SEN) is set to ground \changed{and SL to the supply voltage}, \changed{which precharges the two selected complementary  nanodevices as well as the comparing latch at the same voltage}. 
In the second phase, the sense signal is set to the supply voltage, and the voltages on the   complementary devices are discharged to ground \changed{through SL}. The branch with the lowest resistance discharges faster, and causes its associated inverter output to discharge to ground, which latches the complementary inverter output to the supply voltage.
The two output voltages therefore represent the comparison of the two complementary resistance values.
\changed{In our test chip, the read time is approximately $10~ns$ and is due to the high capacitive load associated with our probe testing setup. Without this high capacitive load, the switching time would  be  determined by the time to resolve the initial metastability of the circuit. This switching time can be as fast as $100ps$ in a scaled technology \citep{zhao2014synchronous}.}

We fabricated a differential memory with 2048  devices, therefore implementing a kilobit memory. Each column of complementary nanodevices features a precharge sense amplifier, and row and columns are accessed through integrated CMOS digital decoders. 
The pads of the dies are not protected for electrostatic discharge, and the dies were tested with commercial 22-pads probe cards. 
 In all the experiments, voltages are set using a home made printed circuit board, and pulses voltages are generated using Keysight B1530A pulse generators.
In the design, the precharge sense amplifiers can  optionally be deactivated and by-passed, which allows  \changed{measuring} the  nanodevices resistance directly  through external precision source monitor units
(Keysight B1517a).


\subsection{Design of In-Memory Binarized Neural Network Based on the Differential Memory Building Block}

\changed{This work aims at} implementing  Binarized Neural Networks in hardware.
In these neural networks, the synaptic weights, as well as the neuronal states, can  take only two values, $+1$ and $-1$, while these parameters assume real values in conventional neural networks.
The equation for neuronal value $A_j$ in an usual neural network \newchanged{is:}
\begin{equation}
    \changed{A_j =  f \left( \sum_i W_{ji}X_i +b_j  \right),}
    \label{eq:activ_real}
\end{equation}
where $X_i$ are the neuron inputs,   $W_{ji}$ the synaptic weights values, \changed{$b_j$ a bias term,} and $f$ \newchanged {the neural activation function, which  introduces non-linearity to the network. Typical examples of activation functions are the sigmoid function, the softmax function, and the hyperbolic tangent function. In Binarized Neural Networks, the activation function is much simpler, as it is substituted by the sign function, as shown in equation: }
\begin{equation}
    A_j = \mathrm{sign} \left( \POPCOUNT_i \left( XNOR \left( W_{ji},X_i \right) \right)-T_j \right).
        \label{eq:activ_BNN}
\end{equation}

In this equation,   $T_j$ is \newchanged{the so called threshold of the neuron, and it is learned during training}. $\POPCOUNT$  is the function that counts the number of ones in a series of bits, and $\mathrm{sign}$ is the sign function. 

The training process of binarized neural networks differs from conventional neural networks.
During training, the weights  assume  real weights \changed{in addition to the binary weights, which are equal to} the sign of the  real weights. 
Training employs the classical error backpropagation equations, with several adaptations. The binarized weights  are used in the equations of both the forward and the backward passes, but the real weights \newchanged{change} as a result of the learning rule  \citep{courbariaux2016binarized}. Additionally, as the activation function of binarized neural networks is the $\mathrm{sign}$ function and \newchanged{it} is not differentiable, \newchanged{we consider the sign function as the first approximation of the hardtanh function 
\begin{equation}
     \mathrm{Hardtanh(x) = Clip(x, -1,1)},
     \label{eq:clip}
\end{equation}
and we use of the derivative of this function as a replacement for the derivative of the sign function in the backward pass.  This replacement is a key element for training BNN successfully.} 
\newchanged{The clip interval in equation~\eqref{eq:clip} is not learned and is chosen to be between -1 and 1 for all neurons. 
Using a larger interval would indeed increase vanishing gradient effect,
%
while using a smaller interval would lead to derivatives higher than one, which can cause exploding gradient effects.
}


\changed{Finally, the Adam optimizer is used to stabilize learning \citep{kingma2014adam}. A technique known as batch-normalization is employed at each layer of the neural network \citep{ioffe2015batch}. Batch-normalization  shifts and scales the neuronal activations over  a batch during the training process. \newchanged{This method is optionally} used in conventional neural networks to accelerate and stabilizing learning. Using this technique becomes essential when training binarized neural networks \newchanged{to reach high accuracies, as it ensures} that neuronal activations utilize both $+1$ and $-1$ value. At inference time, batch-normalization is no longer necessary and the threshold learned by this technique can be used directly as neuronal threshold in equation \eqref{eq:activ_BNN}.}

With this \changed{learning technique}, binarized neural networks function surprisingly well. They can achieve near state of the art performance on image recognition tasks such as CIFAR-10 and ImageNet \citep{lin2017towards}.  
After learning, the real weights serve no more purpose and can be discarded. This makes binarized neural networks exceptional candidates for hardware implementation of neural network inference.  Not only  their memory requirements \newchanged{are} minimal (one bit per neuron and synapse), but their arithmetic \newchanged{is also} vastly simplified.
\newchanged{Multiplication operations of eq.~(\ref{eq:activ_real}) are expensive in terms of area and energy consumption, and they are replaced by} one-bit exclusive NOR (XNOR) operations in eq.~\ref{eq:activ_BNN}. Additionally, the real sums  in eq.~(\ref{eq:activ_real}) are replaced by $\POPCOUNT$ operations, which are equivalent to integer sums with a low bit width.

It is possible to implement ASIC Binarized Neural Networks with solely CMOS \citep{ando2017brein,bankman2018always}.
However, a more \changed{optimal implementation can}  rely on emerging non-volatile memories, and  associate logic and memory as closely as possible. This approach can provide non volatile neural networks, and eliminate the von Neumann bottleneck entirely: 
\changed{ the nanodevices can implement the synaptic weights, while the arithmetic can be done in CMOS.}
Most of the literature \changed{proposing the use of emerging memories as synapses relies} on   \newchanged{an ingenious technique} to perform the multiplications and additions of eq.~\ref{eq:activ_real}, relying on analog electronics: the multiplications are done relying on Ohm's law, and the addition on Kirchoff current law \citep{ambrogio2018equivalent,yu2016binary}.
This analog approach can be transposed \changed{directly} to binarized neural networks
\citep{sun2018fully,sun2018xnor,tang2017binary,yu2018neuro}.
However, binarized neural networks  are inherently digital objects that rely, as previously remarked, on simple logic operation: \changed{XNOR operations and low bit-width sums}. 
Therefore, \changed{here}, we investigate their implementation with purely digital circuitry. This concept has also recently appeared in \citep{natsui2018design,giacomin2019robust} and in our preliminary version of this work \citep{bocquet2018}. Our work is the first one with measurements on a physical memory array, that includes the effect of bit errors.

A first realization is that the XNOR operations can be realized directly within the sense amplifiers. For this, we follow the pioneering works of \citep{zhao2014synchronous}, which shows that precharge sense amplifier can be enriched with any logic operation. In our case, we can add four additional transistors in the discharge branches of a precharge sense amplifier (Fig.~\ref{fig:PCSAs}(b)). These transistors can prevent the discharge and allow implementing the XNOR operation between input voltage $X$ and the value stored in the complementary OxRAM devices in a single operation.

\begin{figure}[h!]
\begin{center}
\includegraphics[width=15cm]{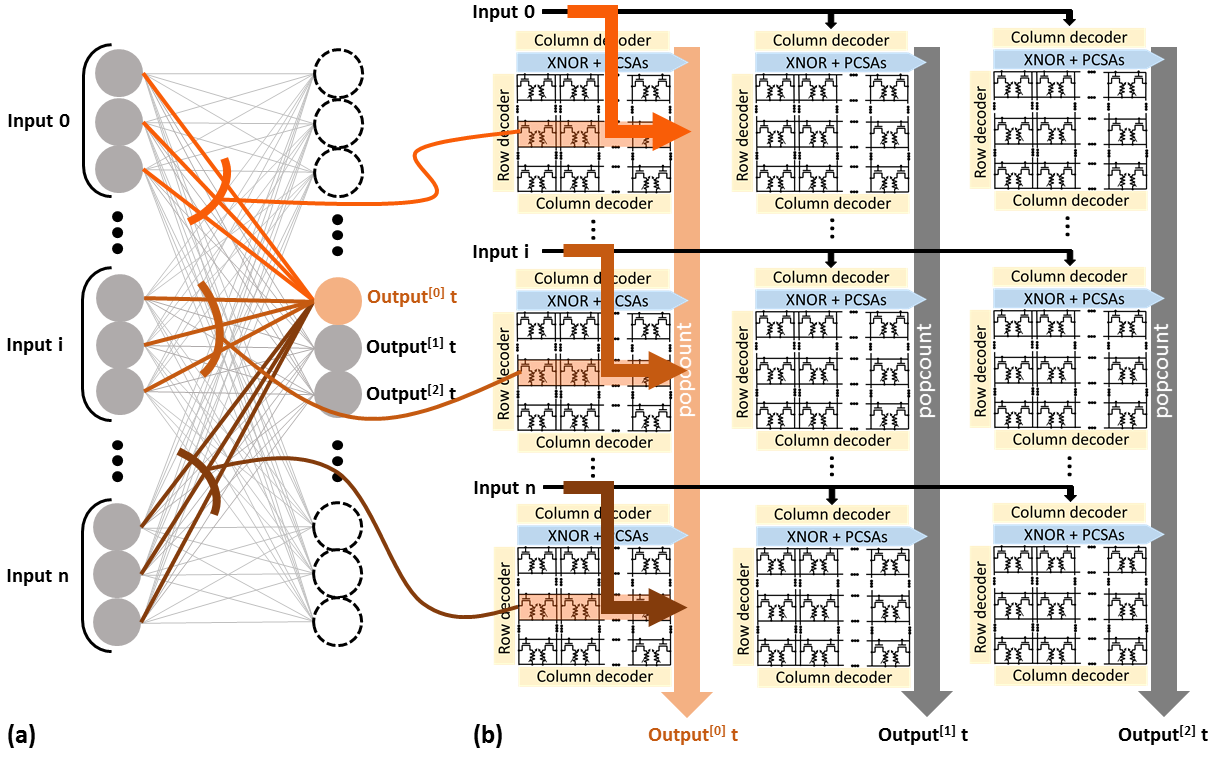}
\end{center}
\caption{Schematization of the full architecture to implement Binarized Neural Network, in the ``parallel to sequential'' configuration. The system assembles  memory block surrounded by logic circuits, and moves minimal data between the blocks. \changed{The architecture is presented with three rows and three columns (i.e. $N=M=3$) of kilobit memory blocks (i.e. $n=32$).}}\label{fig:CMOS_BNN}
\end{figure}

Based on the basic memory array with PCSAs enriched with XNOR, we designed the whole system implementing a Binarized Neural Network. 
The overall architecture is presented in Fig.~\ref{fig:CMOS_BNN}.
It is inspired by the purely CMOS architecture proposed in \citep{ando2017brein},
adapted to the constrains of OxRAM.
\changed{
The design is made of the repetition of basic cells 
organized in a matrix of $N$ by $M$  cells.
These basic cells incorporate a $n\times n$ OxRAM memory block with XNOR-enriched PCSAs and $\POPCOUNT$ logic.
The whole system, which aims at computing the activation of  neurons (equation~\eqref{eq:activ_BNN}), features a degree of reconfigurability to adapt to different neural network topologies:  
it can be used  either in ``parallel to sequential''  or in a ``sequential to parallel'' configuration.}

\changed{
The parallel to sequential configuration (presented in Fig.~\ref{fig:CMOS_BNN}) can deal with layers with up to $n\times N$ input neurons, and up to $n \times M$ output neurons. 
In this situation, at each \newchanged{clock cycle}, the system computes the activations of $M$ output neurons in parallel.
At each clock cycle, each basic-cell reads an entire row of its OxRAM memory array, while perforning the XNOR operation with input neuron values. The results are used to compute the $\POPCOUNT$ operation over a subset of the indices $i$ in equation~\eqref{eq:activ_BNN}, using fully digital five bits counters embedded within the cell.
Additional logic, called ``popcount tree''and only activated in this configuration, computes the full $\POPCOUNT$ value operation over a column by successively adding the five bits-wide partial $\POPCOUNT$ values.
The activation value of the neuron is obtained by subtracting the complete  $\POPCOUNT$ value at the bottom of the column with a threshold value, stored in a separate memory array;  the signed bit of the result gives the activation value.
At the next clock cycle, the next rows in the OxRAM memory arrays are selected, and the activations of the next $M$ neurons are computed.
 }

\changed{
The sequential to parallel configuration (not presented), by contrast, can be chosen to deal with a neural network layer with up to $n^2$ inputs neurons, and up to $NM$ output  neurons.
In this configuration, each basic cell of the system \newchanged{computes} the activation of one neuron $A_j$. The input neurons  $X_i$ are presented sequentially, by subsets of $n$ inputs.
At each clock cycle, the digital circuitry therefore computes only a part of equation~\eqref{eq:activ_BNN}. 
The partial $\POPCOUNT$ is \changed{looped} to the same cell to compute the whole $\POPCOUNT$ sequentially
After the presentations of all inputs, the threshold is subtracted, the binary activation is extracted and equation~\eqref{eq:activ_BNN} has been entirely computed. 
}

This whole system was designed using synthesizable SystemVerilog. 
The memory \changed{blocks} are described in behavioral SystemVerilog. We synthesized the system
using the 130~nanometer design kit used for fabrication, as well as using the design kit of  an advanced commercial 28~nm process for scaling projection.

All simulations reported in the results sections were performed using the Cadence Incisive simulators. 
The estimates for system-level energy consumption were obtained using the Cadence Encounter tool. We used Value Change Dumps (VCD) files extracted from simulations of practical tasks so that the obtained energy values reflect the operation of the system realistically.

\section{Results}

\subsection{Differential Memory Allows Memory Operation at Reduced Bit Error Rate}

\begin{figure}[h!]
\begin{center}
\includegraphics[height=45mm]{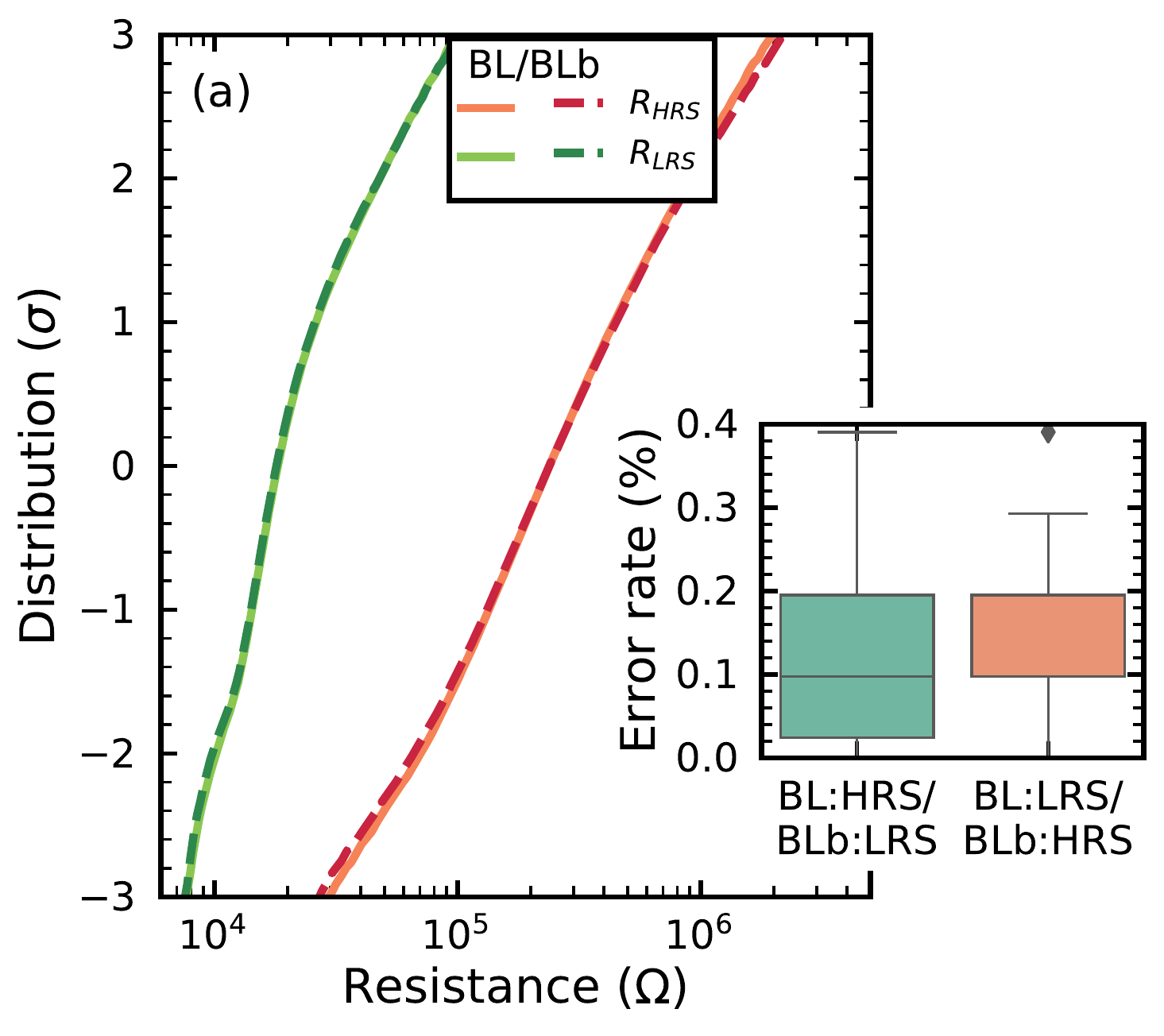}\hfill
\includegraphics[height=45mm]{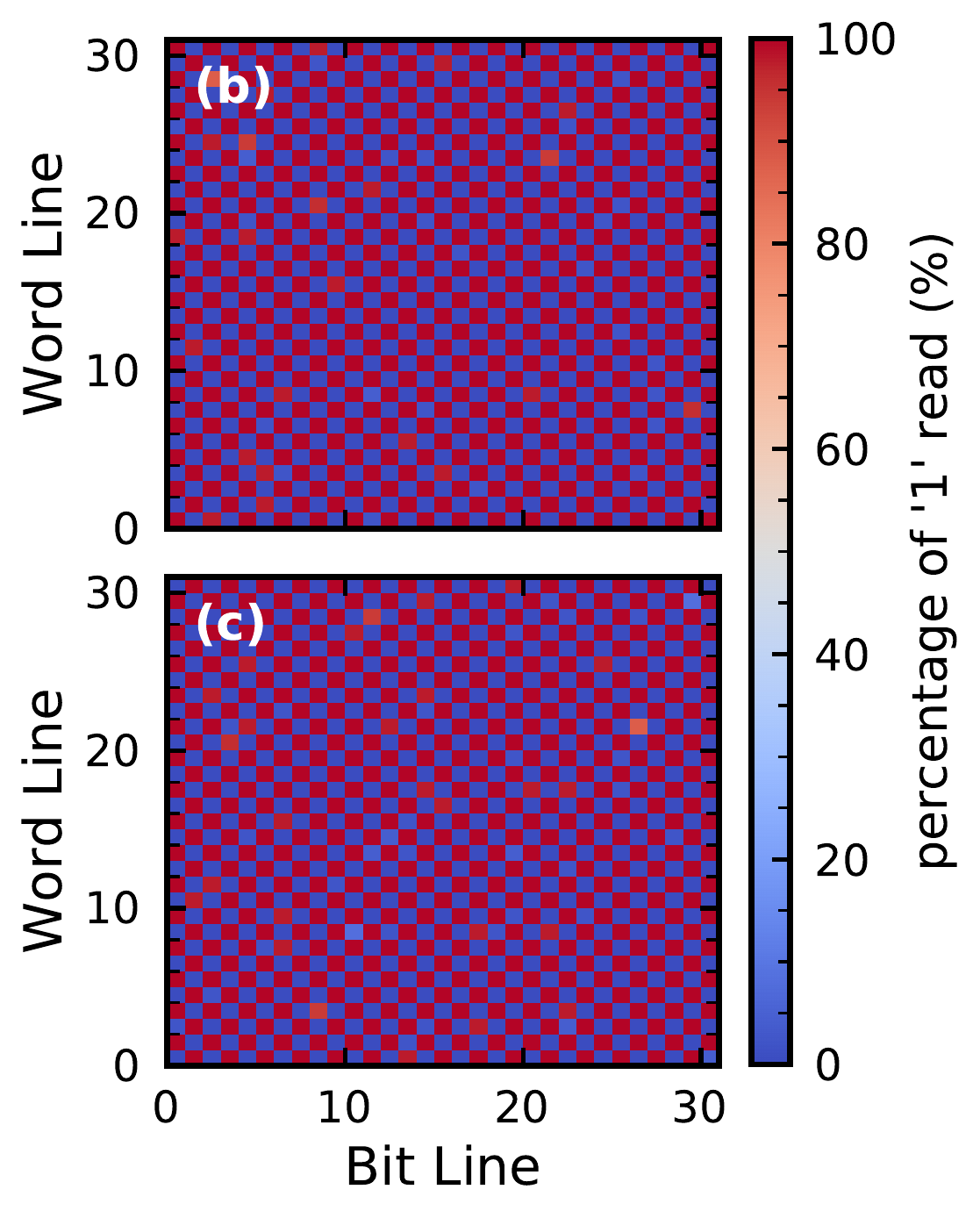}\hfill
\includegraphics[height=45mm]{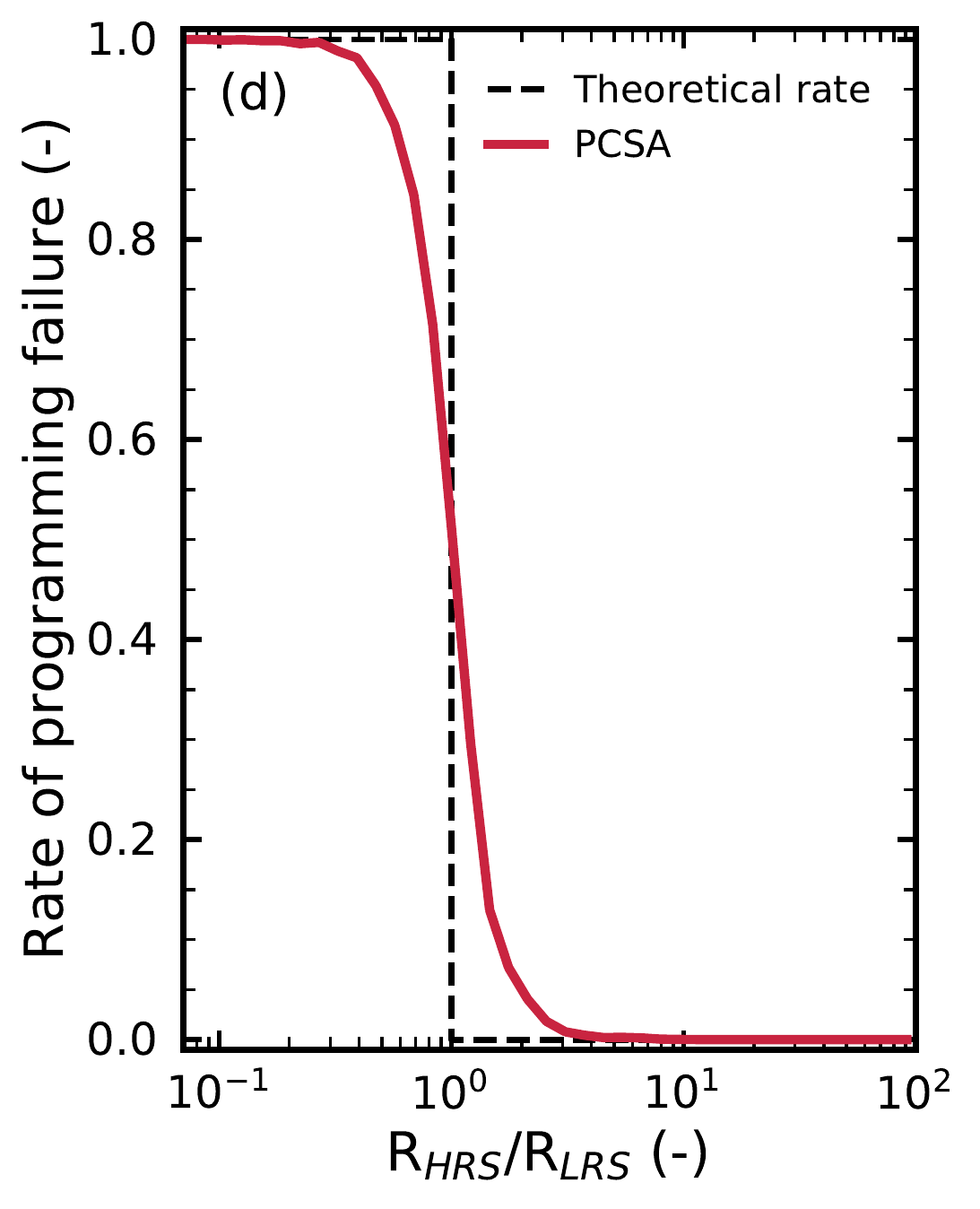}\hfill
\includegraphics[height=45mm]{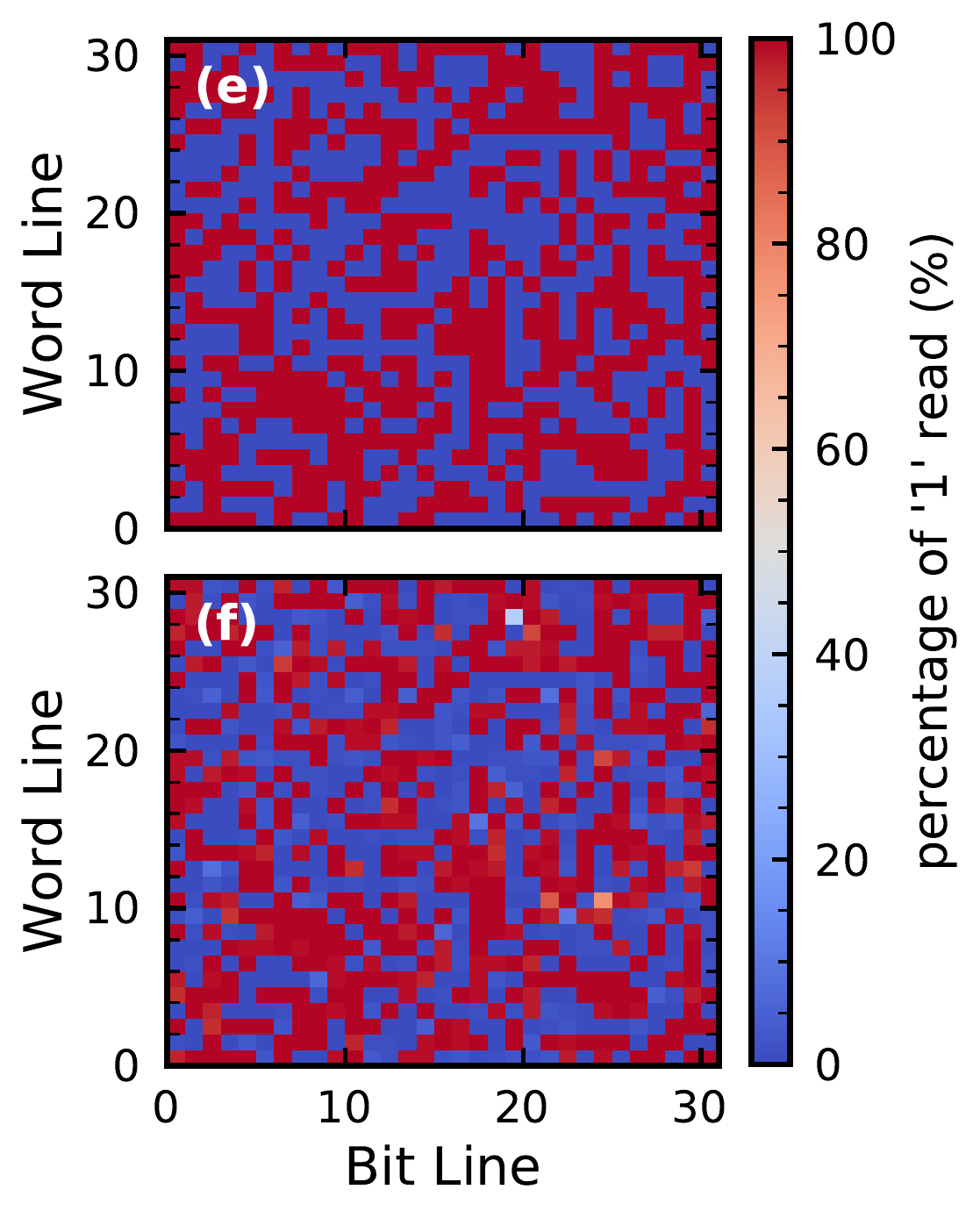}
\end{center}
\caption{(a) Distribution of the LRS and the HRS of the OxRAM devices in an array programmed with a checkerboard pattern.  
RESET voltage of  $2.5V$,  SET current of  $55 \mu A$ and  programming time of $1\mu s$.
(b-c) \changed{Proportion of $1$ values read by the onchip precharge sense amplifier, over} 100 whole-array programming of a memory array, for the  two complementary checkerboards configuration. 
(d) Rate of programming failure indicated of the precharge sense amplifier circuits as a function of the ratio between HRS and LRS resistance (measured by a sense measure unit), in the same configuration as (a-c). \changed{(e-f) Proportion of $1$ values read by the onchip precharge sense amplifier, over 100 whole-array programming of a memory array, for the  last layer of a binarized neural network trained on MNIST (details in body text). }
}\label{fig:errors_intro}
\end{figure}

This section first presents the electrical characterization results of the differential OxRAM arrays.
We program the array with  checkerboard-type data, alternating zero and one bits, using programming times of one microsecond. For programming devices in HRS (RESET operation), the access transistor is fully opened and a  reset voltage of $2.5V$ is used. 
For programming devices in LRS (SET operation), the gate voltage of the access transistor is chosen to ensure a compliance current of $55 \mu A$.
Fig.~\ref{fig:errors_intro}(a) shows the statistical distribution of the LRS and HRS of \changed{the cells}, based on 100 programmings of the full array.  
This graph is using a standard representation in the memory field, where the \textit{y} axis is expressed as number of standard deviations of the distribution \citep{ly2018role}.
The Figure superimposes distributions of left (BL) and right (BLb) columns of the array, and \changed{no significant difference is seen between BL and BLb devices}.
The LRS and HRS distributions are clearly separate but overlap at a value of three standard deviations, \changed{which makes} bit errors possible. If a 1T1R structure was used, a bit error rate of $0.012$ ($1.2\%$) would be seen with this distribution. \changed{By contrast,} at the output of the precharge sense amplifiers, a bit error rate of $0.002$ ($0.2\%$) is seen, \changed{giving} a first suggestion of the benefits of the 2T2R approach.
Fig.~\ref{fig:errors_intro}(b) and \ref{fig:errors_intro}(c) show the  mean error (using the 2T2R configuration) on the whole array, for the two types of checkerboards.  We see that all devices can be programmed in HRS and LRS. A few devices have increased bit error rate. This graph highlights the existence of both cycle to cycle and device to device variability, and the absence of \changed{``dead'' cells}. 

We now validate in detail the functionality of the precharge sense amplifiers. The precise resistance of devices is first measured by deactivating the precharge sense amplifiers, and using the external source monitor units. Then, the precharge sense amplifiers  are reactivated and a sense operation is performed. Fig.~\ref{fig:errors_intro}(d) plots the mean measurement of the sense amplifiers as a function of the ratio between the two resistances that are being compared, superimposed with the ideal behavior of a sense amplifier. 
The sense amplifiers show excellent functionality, but can make mistakes if the two resistances differ by less than a factor five.
\changed{Finally, Figs.~\ref{fig:errors_intro}(e-f) repeat the experiments of \ref{fig:errors_intro}(b-c) in a more realistic situation and on a different die. 
We trained a memory array 100 times with weights corresponding to the last layer of a binarized neural network trained on the MNIST task of handwritten digit recognition. 
As in the checkerboard case,  no dead cell is seen, and a similar degree of cycle-to-cycle and device-to-device variation is seen.}

\begin{figure}[h!]
\begin{center}
\includegraphics[width=10cm]{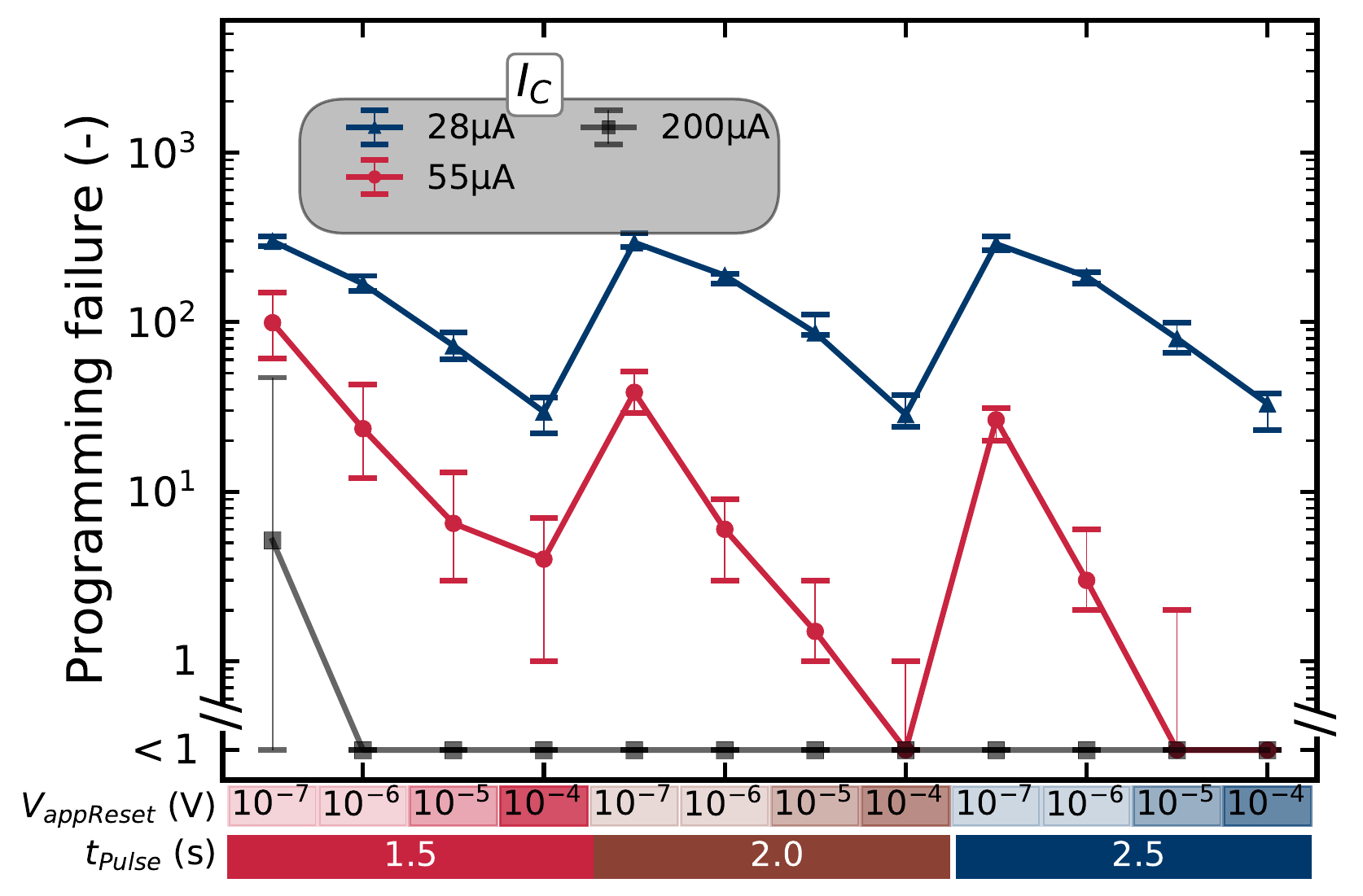}
\end{center}
\caption{
\changed{Number of  errors} for different programming
conditions, as measured by the precharge sense amplifier, for 2T2R configuration  on a kilobit
memory array.
\newchanged{ The ``$<1$'' label means that no errors were detected. The error bars present the minimum and maximum number of detected errors, over five repetitions of the experiments. }
}\label{fig:prog_cond}
\end{figure}

The programming rates are strongly dependent on the programming conditions. Fig.~\ref{fig:prog_cond} shows the mean number of  incorrect bits on a whole array for various combination of programming times \changed{(from \newchanged{$0.1\mu s$} to $100\mu s$), RESET voltage (between $1.5$ and $2.5$ Volts), and SET compliance current (between $28$ and $200 \mu A$)}. We \changed{observe} that the bit error rate depends extensively on \changed{these} three programming parameters, the SET compliance current having the most significant impact.

\begin{figure}[h!]
\begin{center}
\includegraphics[width=17cm]{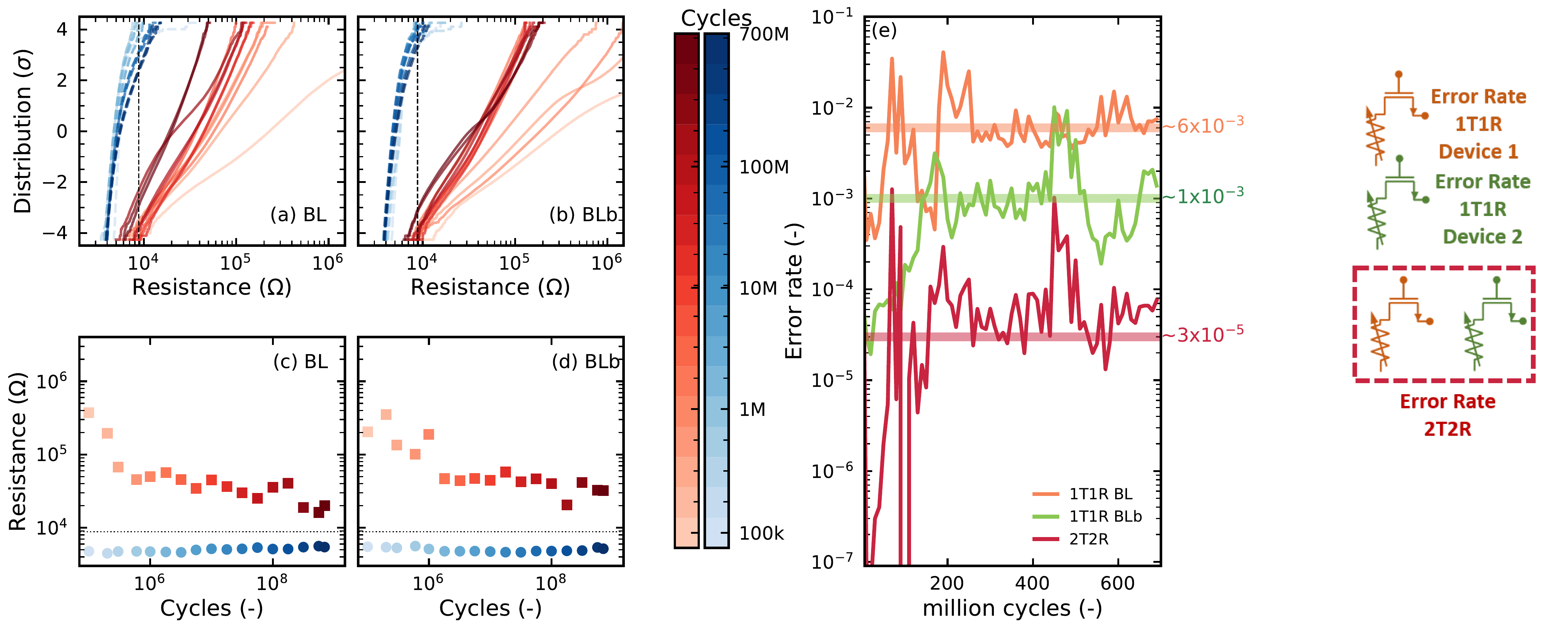}

\end{center}
\caption{
\changed{
(a-b) Distribution of the resistance values, (c-d)  mean resistance value and (e) mean bit error rate over 10 million cycles measured by the precharge sense amplifier, in the 2T2R configuration, as
function of the number of cycles that a device has been programmed. 
RESET voltage of  $2.5V$,  SET current of  $200 \mu A$ and  programming time of $1\mu s$.
}
}\label{fig:2T2R_vs1T1Raging}
\end{figure}

\begin{figure}[h!]
\begin{center}
\includegraphics[width=17cm]{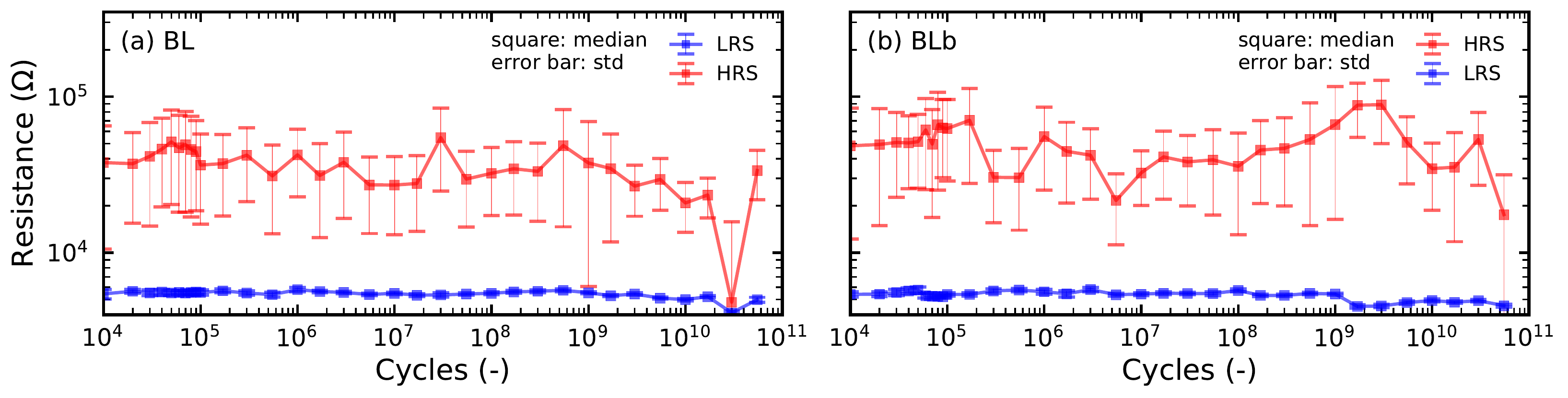}
\end{center}
\caption{
\newchanged{
(a-b)  Mean resistance value of the BL and BLb device over 10 thousand cycles for  measurements of a device pair over $5\times 10^{10}$ cycles. RESET voltage of  $1.5V$,  SET current of  $200 \mu A$ and  programming time of $1\mu s$.
}
}\label{fig:Fig7}
\end{figure}

In Fig.~\ref{fig:2T2R_vs1T1Raging}, we look more precisely at the effects of cycle to cycle device variability and device aging. A device and its complementary device were programmed $700$ million cycles.
Figs.~\ref{fig:2T2R_vs1T1Raging}(a) and \ref{fig:2T2R_vs1T1Raging}(b) show the distribution of LRS and HRS of the device under test and its complementary device, after different number \changed{of} cycles ranging from the first one \changed{to} the last one. 
\newchanged{We can observe that}
when the devices are cycled, LRS and HRS distributions become less separated and start to overlap at lower number of standard deviations. 
This translates directly on the mean resistance of the devices in HRS and LRS (Figs.~\ref{fig:2T2R_vs1T1Raging}(c) and \ref{fig:2T2R_vs1T1Raging}(d)), which become closer when the device ages.
More importantly, the aging process \newchanged{impacts} the device bit error rate (\newchanged{Figs.~}\ref{fig:2T2R_vs1T1Raging}(e)): the bit error rate of the device and its complementary device increase of several orders of magnitudes over the lifetime of the device. 
The same \changed{effect} is seen on the bit error rate resulting from the precharge sense amplifier (2T2R), but it remains at much lower level: 
while the 1T1R bit error rate \newchanged{goes} above $10^{-3}$ after a few million cycles, the 2T2R remains below this value over the $700$ million cycles.
This result highlights that the concept of cyclability depends  on the acceptable bit error rate, and that the cyclability at constant bit error rate can be considerably extended when using the 2T2R structure.
\changed{It should also be highlighted that the cyclability depends tremendously on the programming condition. \newchanged{Fig.~\ref{fig:Fig7}(a-b))} shows endurance measurements  with a reset voltage of $1.5V$ (all other programming conditions are identical to Fig.~\ref{fig:2T2R_vs1T1Raging}(a-e)). \newchanged{We can see} that the device experiences no degradation along more than  ten billion cycles. \newchanged{Over} that time, the 2T2R bit error rates remains below $10^{-4}$.}

\begin{figure}[h!]
\begin{center}
\includegraphics[width=17cm]{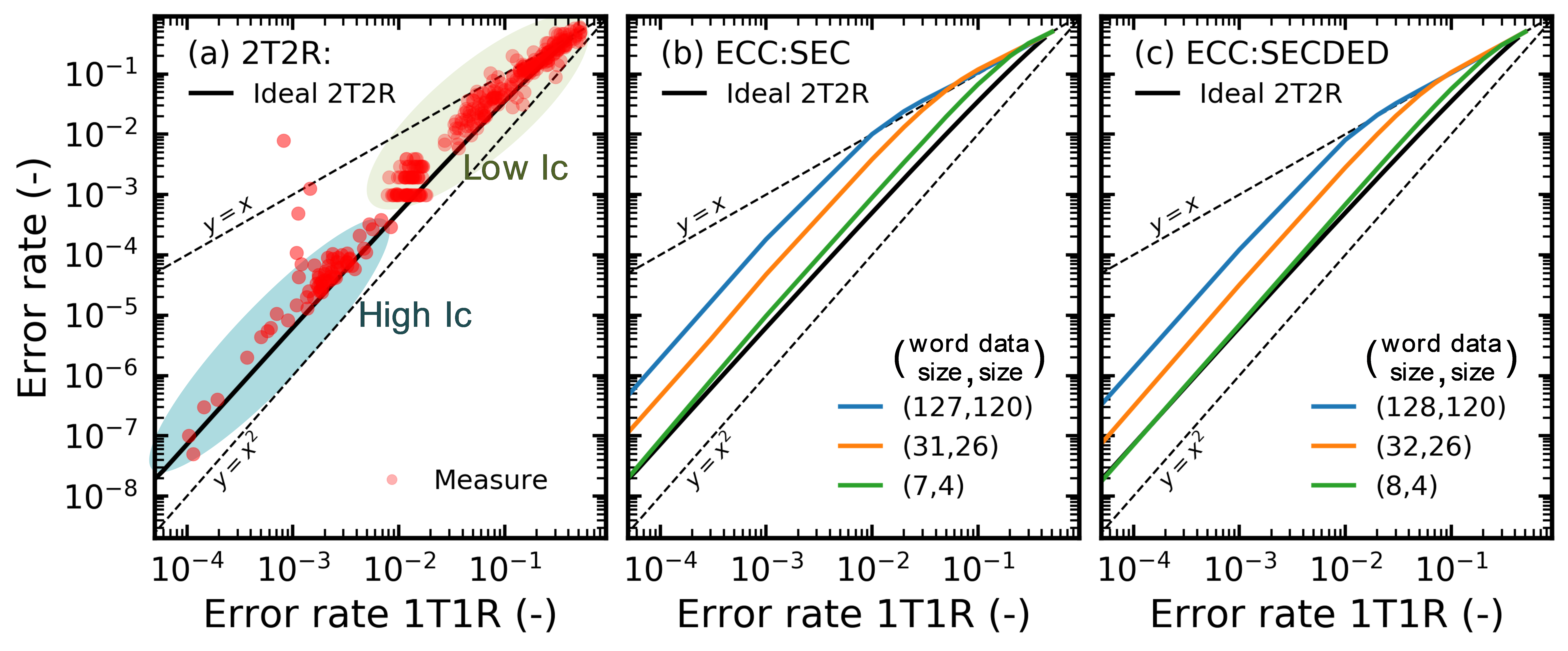}
\end{center}
\caption{
Experimental bit error rate of the 2T2R array, measured by the precharge sense amplifiers, as a function of the bit error rate obtained individual (1T1R) RRAM devices in the same programming conditions.
The detailed methodology for obtaining this graph is presented in the body text.
Bit error rate obtained with (b) Single Error Correcting  (SEC) and (c) Single Error Correcting  Double Error Detection (SECDED) ECC as a function of the error rate on the individual devices.
}
\label{fig:ECC}
\end{figure}

We now aim at quantifying and benchmarking more precisely the benefits of the 2T2R structure.
We performed extensive characterization of \changed{bit error rates} on the memory array in various regimes. 
Fig.~\ref{fig:ECC}(a) presents different experiments where the 2T2R bit error rate 
is plotted as a function of the bit error rate that would be obtained using using a single device programmed in the same conditions. \changed{The different points are obtained by varying the compliance current Ic during SET operations, and the} graph associates two type of experiments: 

\begin{itemize}
    \item The points marked as ``Low Ic'' are obtained using whole array measurement where devices are programmed with low SET compliance current to ensure high error rate. 
    \changed{Each device  in the memory array is programmed once (following the checkerboard configuration), and all synaptic weights are read  using the on-chip precharge sense amplifiers. The plotted bit error rate is the proportion of weights for which the read weight differs from the  weight value targeted by the programming operation.}
    \item \changed{The points marked as ``High Ic'' are obtained by  measurements on a single device pair. A single 2T2R structure in the array is  programmed ten million times by alternating programming to $+1$ and $-1$ values. The value programmed in the 2T2R structure is  read using an on-chip precharge sense amplifier after each programming operation. The plotted bit error rate is the proportion  of read operation for which the read weight differs from the  targeted value.}
\end{itemize}

\newchanged{We can see} that the 2T2R bit error rate is always lower than the 1T1R one. The difference is larger for lower bit error rate, and reaches four order of magnitudes for a 2T2R bit error rate of $10^{-8}$.
The black line presents calculation where the precharge sense amplifier is supposed to be ideal (\textit{i.e.} to follow the idealized dotted characteristics of Fig.~\ref{fig:errors_intro}(c)).

To interpret the results of the 2T2R approach with more perspective, we benchmark them with standard error correcting codes.
Figs.~\ref{fig:ECC}(b) and \ref{fig:ECC}(c) show the benefits of two codes, using the same plotting format as   Figs.~\ref{fig:ECC}(a):
a Single Error Correction (SEC)  and a Single Error Correction Double Error Detection (SECDED) code, presented with different degrees of redundancy.
These  simple codes, formally known as Hamming and extended Hamming codes, are widely used in the memory field.
Interestingly, we see that the benefit of these codes are very similar to the benefit of our 2T2R approach with ideal sense amplifier, at equivalent memory redundancy (\textit{e.g.} SECDED(8,4)), although our approach uses no decoding circuit and the equivalent of error correction is performed directly within the sense amplifier. By contrast, ECCs  can also reduce bit errors, to a lesser extent, using less redundancy\changed{, but} the required decoding circuits \changed{utilize}  hundreds to thousands of logic gates \citep{gregori2003chip}. 
In a context where logic and memory are tightly integrated, these decoding circuits would need to be repeated many times, \changed{ and} as their logic is much more complicated than the one of binarized neural networks, they would be the dominant source of computation and energy consumption.
ECC circuits are also incompatible with the idea of integrating XNOR operations within the sense amplifiers, and cause important read latency.
\\


\subsection{Do All Errors Need to Be Corrected?}
\label{subsec:DoAllErrorsNeedBeCorrected}

\begin{figure}[h!]
\begin{center}
\includegraphics[width=10cm]{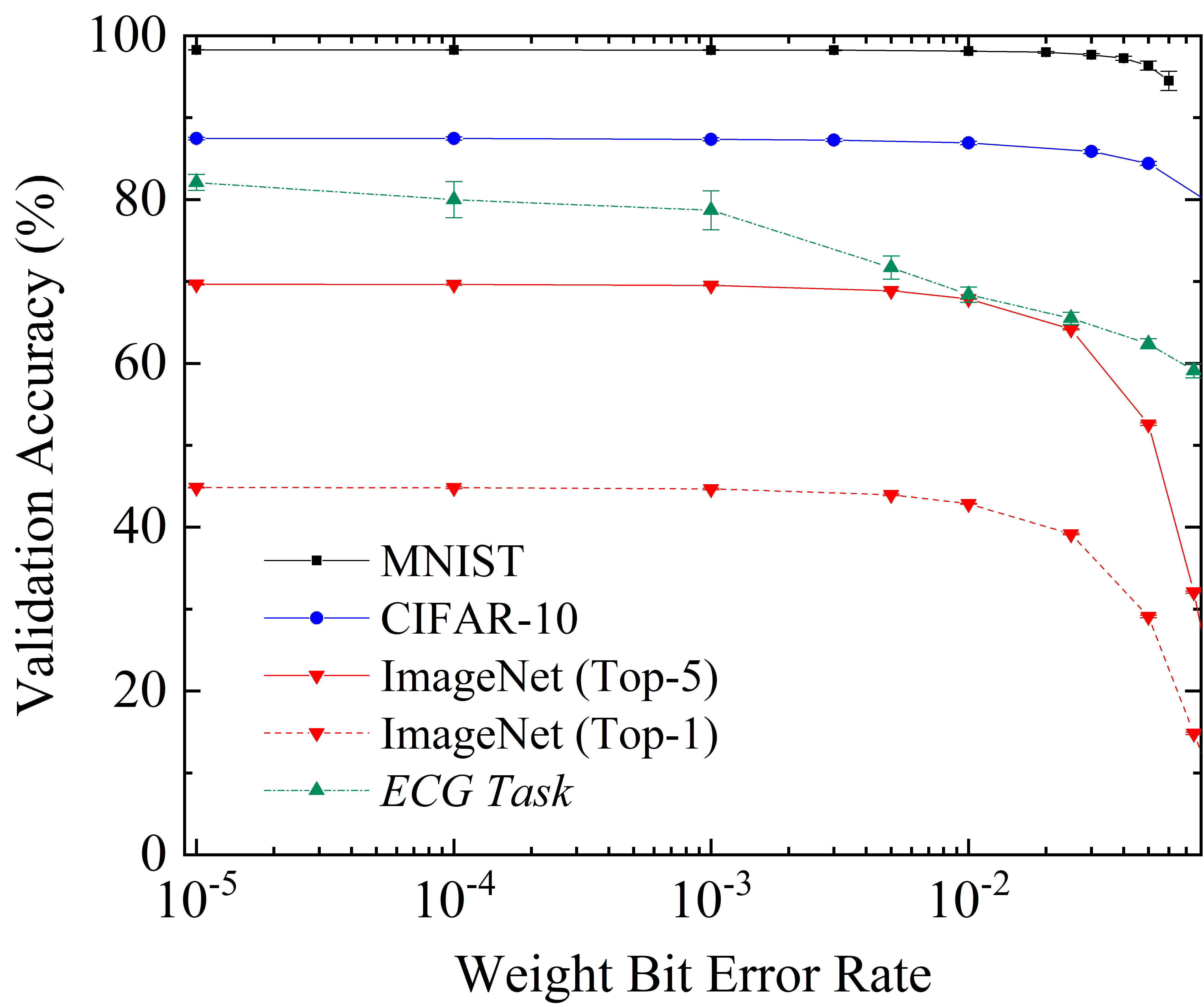}

\end{center}
\caption{Recognition rate on the validation dataset of the fully connected neural network for MNIST, the convolutional neural network for CIFAR10, and AlexNet for ImageNet (Top-5 and Top-1) accuracies  and the ECG analysis task, as a function of the bit error rate over the weights during inference.  
	    Each experiment was repeated five times, the mean recognition rate is presented. 
	    \changed{Error bars represent one standard deviation.}
	    }\label{fig:bit_errorsBNN}
\end{figure}

Based on the results of \changed{the} electrical measurements, and before \changed{discussing} the whole system, it is important to know  the OxRAM bit error rate  \changed{levels that can be tolerated} for applications.
To answer this question, we performed simulations of binarized neural networks on four different tasks:

\begin{itemize}

    \item  MNIST handwritten digit classification \citep{lecun1998gradient}, the canonical task of machine learning. We use a  fully connected neural network with two 1024-neurons hidden layers.
    
    \item The CIFAR-10 image recognition task \citep{krizhevsky2009learning}, which consists in  recognizing $32 \times 32$ color images spread between ten categories of  vehicles and animals. We use a deep convolutional network with six convolutional layers using kernels of  $3 \times 3$ and  a stride of one, followed by \changed{three  fully} connected layers.  
    
    \item The ImageNet  recognition task,  which consists in  recognizing  \changed{$224 \times 224$} color images out of 1000 classes. This task is considerably more difficult than MNIST and CIFAR-10. We use the historic  AlexNet deep convolutional neural network  \citep{krizhevsky2012imagenet}.
    
    \item  A medical task involving the analysis of electrocardiography (ECG) signals: automatic detection of electrode misplacement.
    This binary classification challenge takes as input the ECG signals of twelve electrodes. The experimental trial \changed{data} are sampled at $250Hz$    and have a duration of three seconds each.
    To solve this task, we \changed{employ} a convolutional neural network
    composed  of five convolutional layers and two fully-connected layers. The convolutional kernel (sliding
    window) sizes \changed{are} decreasing from 13 to 5 in each subsequent layer.
    Each convolutional layer produce 64 filters 
    detecting different features of the signal.
\end{itemize}

Fully binarized neural networks were trained on these tasks on Nvidia Tesla GPUs using Python and the PyTorch deep learning framework. Once the neural networks \changed{were} trained, we ran them on the datasets validation sets, artificially introducing errors in the neural networks weights (meaning some $+1$ weights are replaced by $-1$ weights, and reciprocally).
\changed{Using this technique}, we \changed{can} emulate the impact of OxRAM bit errors. 
Fig.~\ref{fig:bit_errorsBNN} shows the resulting validation accuracy as a function of the introduced bit error rate for the four considered tasks. In the case of ImageNet, both the Top-1 (proportion of validation images where the right label is the top choice of the neural network) and the Top-5 (proportion of validation images where the right label is within the top five choices of the neural network).

On the three vision task (MNIST, CIFAR-10, ImageNet), we see that extremely high levels of bit errors can be tolerated:
up to a bit error rate of $10^{-4}$, the network performs as well as with no errors.
Minimal performance reduction starts to be seen with bit error rates of  $10^{-3}$ (the Top-5 accuracy on ImageNet is degraded from $69.7\%$ to  $69.5\%$). At bit error rates of $0.01$, the performance reduction becomes significant. 
The reduction is more substantial for harder tasks: MNIST accuracy is only degraded from $98.3\%$ to $98.1\%$, CIFAR-10 accuracy is degraded from $87.5\%$ to $86.9\%$, while ImageNet Top-5 accuracy is degraded from  $69.7\%$ to $67.9\%$.

The ECG tasks also shows extremely high bit error tolerance, but bit errors have an effect more rapidly than in the vision tasks. At a bit error rate of $10^{-3}$, the validation accuracy is reduced from $82.1\%$ to $78.7\%$, and at a bit error rate of $0.01$ to $68.4\%$. This difference between vision and ECG tasks probably originates in the fact that ECG signals carry a lot less redundant information than images. Nevertheless, we see that even for ECG tasks  high bit errors rates can be accepted with regards to the standards of conventional digital electronics.


\section{Discussion}

\subsection{Projection at the System Level}

\subsubsection{Impact of In-Memory Computation}

We now use all the paper results to discuss the potentials of our approach.
Based on our ASIC design, using the energy evaluation technique described at the end of the Methods section, we find that our system would consume $25~nJ$ to recognize one handwritten digit, using a fully connected neural networks with two hidden layers of 1024 neurons.
This is considerably less than processor based options.
\changed{For example,} \citep{lane2016deepx} analyses the energy consumption of inference on CPUs and GPUs:  \changed{operating} a fully connected neural network with two hidden layers of 1000 neurons requires $7$ to $100$ millijoules on a low power CPU (from Nvidia Tegra K1 or Qualcomm Snapdragon 800 systems on chip), and $1.3$ millijoules on a low power GPU (Nvidia Tegra K1).

These results are not surprising due to the considerable overhead for accessing memory in modern computers. For example, \citep{pedram2017dark} shows that accessing data in a static RAM cache consumes around fifty times more energy than the integer addition of this data. If the data is stored in the external dynamic RAM, the ratio  is \changed{increased} to more than 3000. 
Binarized Neural Networks require minimal arithmetic: no multiplication, and only integer addition with a low \changed{bit width}. When operating a Binarized Neural Networks on a CPU or GPU, the almost entirely of the energy is used to move data, and the inherent topology of the neural network is not exploited to reduce data movement. Switching to in-memory or near-memory computing approaches has therefore the potential to reduce energy consumption drastically for such tasks. This is especially true as, in inference hardware, synaptic weights are static and can be programmed to memory only \newchanged{once} if the circuit does not need to change function.

\subsubsection{Impact of Binarization}

\begin{figure}[h!]
\begin{center}
\includegraphics[width=8.5cm]{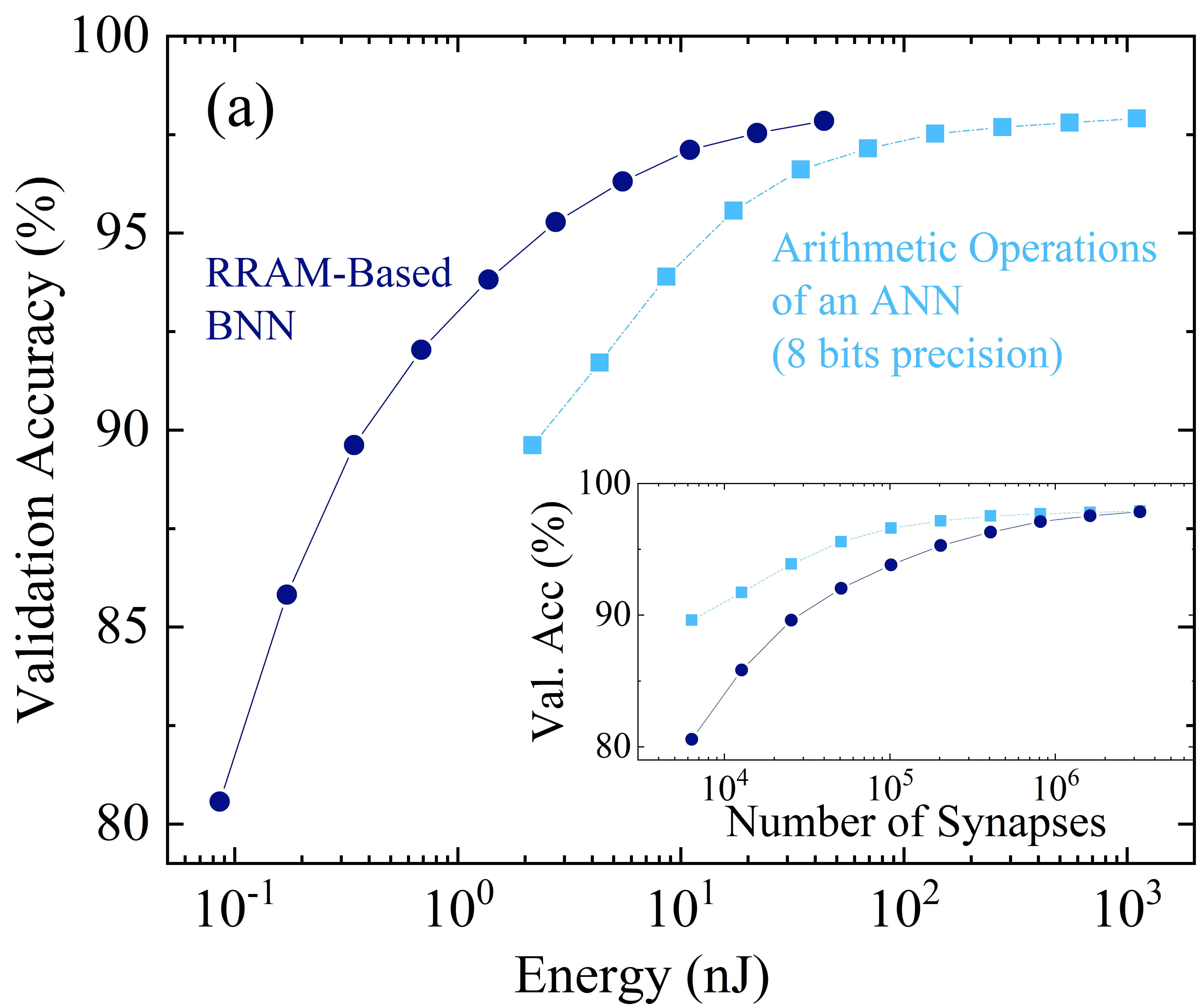}
\includegraphics[width=8.5cm]{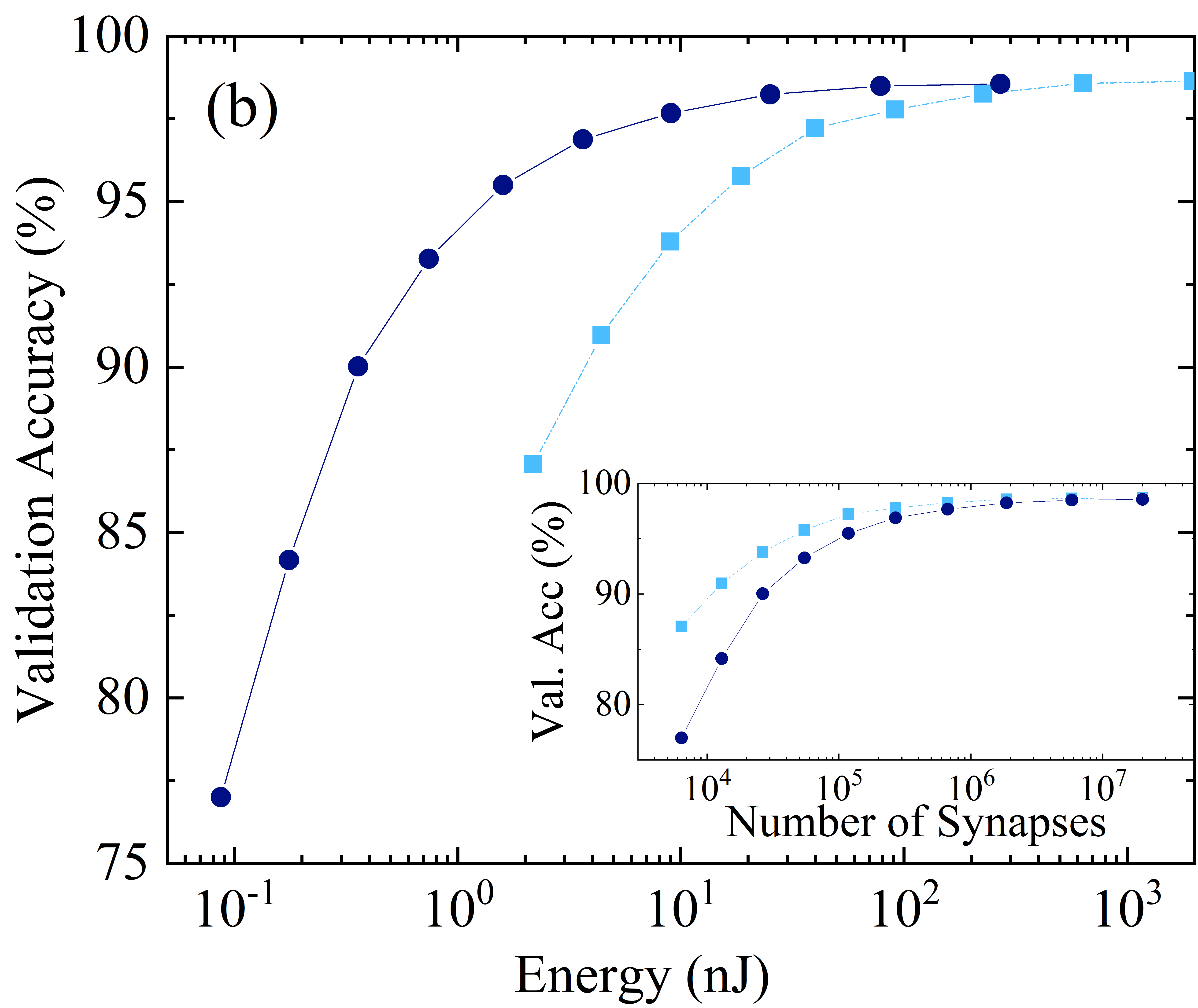}
\end{center}
\caption{\newchanged{Dark blue} circles: MNIST validation accuracy as a function of the inference energy of our Binarized Neural Network hardware design. \newchanged{Light blue square}: same, as function of the energy used for arithmetic operation in a real valued neural networks employing eight bits fixed point arithmetic. The different points are obtained by varying the number of hidden neurons in (a) a one hidden layer neural network and (b) a two hidden layers neural network. Insets: number of synapses in each situations.
}
\label{fig:ANNvsBNN}
\end{figure}

We now look specifically to the benefits of relying on Binarized Neural Networks rather than real-valued digital ones.
Binarized Neural Networks feature considerably simpler architecture than conventional neural network, but also require an increased number of neurons and synapses to achieve equivalent accuracy. It is therefore essential to confront the binarized and real-values approaches.

Most digital ASIC implementations of neural networks inference function with eight-bit fixed point arithmetic, the most famous example being the tensor processing units developed by Google \citep{jouppi2017datacenter}. 
At this precision, no degradation is usually seen for inference with regards to 32 and 64~bits floating point arithmetic.

To investigate the benefits of Binarized Neural Networks, Fig.~\ref{fig:ANNvsBNN} looks at the energy consumption for inference over a single MNIST digits. We consider two architectures: a neural network with a single hidden layer (Fig.~\ref{fig:ANNvsBNN}(a)) and another one with two hidden layers (Fig.~\ref{fig:ANNvsBNN}(b)), and we vary the number of hidden neurons. Figs.~\ref{fig:ANNvsBNN}(a) and  Figs.~\ref{fig:ANNvsBNN}(b) plot on the \textit{x} axis the estimated energy consumption of a Binarized Neural Networks using our architecture based on the flow presented in the Methods section. It also plots the energy required for the arithmetic operations (sum and product) of a eight bit fixed point regular neural network, neglecting \changed{the} overhead that is considered for the Binarized Neural Network. For both types of networks, the \textit{y} axis shows the resulting accuracy on the MNIST task.
We see that at equivalent precision, the Binarized Neural Network always consume  less energy than the arithmetic operations of the real-valued one. It is remarkable that the energy benefit depends  significantly on the targeted accuracy, and should therefore be investigated in a case by case basis. The highest energy benefits, a little less than a factor ten, are seen at lower targeted precision.

Binarized Neural Networks have other benefits with regards to real valued digital networks: if the weights are stored in RRAM, the programming energy is reduced due to the lower memory requirements of Binarized Neural Networks.
The area of the overall circuit is also expected to be reduced due to the absence of multipliers, which \changed{are}  high area circuit.

\subsubsection{Comparison with Analog Approaches}

As mentioned in the introduction, a widely studied approach for implementing neural networks with RRAM is to rely on an analog electronics strategy, where Ohm's law is exploited for implementing multiplications, and Kirchoff's current law for implementing additions \citep{ambrogio2018equivalent,prezioso2015training,serb2016unsupervised,li2018efficient,wang2018fully}.
\newchanged{The digital approach presented in this paper cannot be straightforwardly compared to the analog approach:
the detailed performance of the analog approach  depends tremendously on its implementation details, device specifics and size of the neural networks.} 
Nevertheless, several points can be \newchanged{raised}.

First, the programming of the devices is much simpler in our approach than in the analog one: one only needs to program a device and its complementary one in LRS and HRS, which can be achieved by two programming pulses.  It is not necessary to verify the programming operation, as the neural network has inherent bit error tolerance. Programming RRAM for analog operation is \changed{a} more challenging task, and usually requires a sequence of multiple pulses \citep{prezioso2015training}, which leads to higher programming energy and device aging.

For the neural network operation, the analog approach and ours function differently. Our approach reads synaptic values using the sense amplifier, which is a highly energy efficient and fast circuit that can operate at hundreds of picoseconds in advanced CMOS nodes \citep{zhao2014synchronous}. This sense amplifier inherently produces the multiplication operation, and then the addition needs to be performed \changed{using} low bit-width digital integer addition circuit. The ensemble of a read operation and the corresponding addition typically consumes fourteen femtojoules in our estimates in advanced node. In the analog approach, \changed{the read operation} is performed by applying a voltage pulse, and inherently produces the multiplication though Ohm's law, but also the addition though Kirchoff law. \changed{This approach is} attractive, but in the other hand requires the use of CMOS analog overhead circuitry such as operational amplifier, which can bring \changed{high} energy and area overhead. Which approach is the most energy efficient between ours and the analog one will probably depend tremendously \newchanged{on} memory size, application and targeted accuracy.

Another advantage of the digital approach is that it is much simpler to design, test and verify, as it relies on all standard VLSI design tools. On the other hand, an advantage of the analog approach is that it may, for small memory size, function without access transistors, resulting in higher memory densities \citep{prezioso2015training}.


\subsubsection{Impact in Terms of Programming Energy and Device Aging}


\begin{table}[htbp]
\caption{RRAM Properties with Different Programming Conditions
}
\begin{center}
\begin{tabular}{|m{4cm}|m{3.8cm}|m{4.0cm}|m{3.8cm}|}
\hline
\newchanged{\textbf{Programming condition}} &
\newchanged{\textbf{Strong (Fig. 6)}} & \newchanged{\textbf{Optimized  endurance (Fig. 7)}} 
& \newchanged{\textbf{Optimized  programming energy}} \\
\hline
\newchanged{SET compliance current } &
\newchanged{$200\mu A$} & \newchanged{$200\mu A$} & \newchanged{$200\mu A$}\\
\hline
\newchanged{RESET voltage} &
\newchanged{$2.5V$} &
\newchanged{$1.5V$} &
\newchanged{$2V$} \\
\hline
\newchanged{Programming time} & 
\newchanged{$1\mu s$}& \newchanged{$1\,\mu s$} & \newchanged{$100\, ns$}\\
\hline
\newchanged{2T2R Bit error rate (before aging)} &
\newchanged{$< 10^{-7}$} & \newchanged{$<10^{-4}$} & \newchanged{$< 10^{-5}$}\\
\hline

\newchanged{Programming energy } &
\newchanged{$200\sim 400\,pJ$} & \newchanged{$150 \sim 400\,pJ$} & \newchanged{$20 \sim 30\,pJ$} \\
\newchanged{(SET/RESET)} & & &\\
\hline
\newchanged{Cyclability (number of cycles at BER $<10^{-3}$) }&
\newchanged{$>10^{8}$} &
\newchanged{$>10^{10}$} & 
\newchanged{$>10^{8}$}\\
\hline
\end{tabular}
\end{center}
\label{tab:programming_conditions}
\end{table}

A last comment is that the bit error tolerance of binarized neural networks can have considerable benefits at the system level.
Table~\ref{tab:programming_conditions} summarizes the measured properties of RRAM cells in different programming conditions, \newchanged{chosen out of the ones presented in Fig.~\ref{fig:prog_cond}.
We consider only the conditions with bit error rates below $10^{-3}$ (i.e.  corresponding to a ``$<1$'' data point in Fig.~\ref{fig:prog_cond}), as this constrains makes them appropriate for use for all tasks considered in section~\ref{subsec:DoAllErrorsNeedBeCorrected}.
The ``strong'' programming conditions are the ones presented in Fig.~\ref{fig:2T2R_vs1T1Raging}. They feature  low bit error rate before aging, but  high programming energy. 
The other two columns correspond to two optimized choices.
The conditions optimized for programming energy are the conditions of Fig.~\ref{fig:prog_cond} with bit error rates below $10^{-3}$  and the lowest programming energy. They use a lower RESET voltage ($2.0V$) than the strong conditions, and shorter programming time ($100ns$). The cyclability of the device 
-- defined as the number of cycles a cell can be programmed while retaining a bit error rate below $10^{-3}$ --
remains comparable to the strong programming conditions.
The conditions optimized for endurance are by contrast the  conditions of Fig.~\ref{fig:prog_cond} with bit error rate below $10^{-3}$  and the highest cyclability: more than $10^{10}$ cycles, as already evidenced in Fig.~\ref{fig:Fig7}. These conditions use  a low RESET voltage $1.5V$  but require a programming time of $1\mu s$.
}


\subsection{Conclusion}

\changed{
This work proposes} an architecture for implementing binarized neural networks with RRAMs, \changed{and 
 incorporates} several biological-plausible ideas:
\begin{itemize}
    \item Fully co-locating logic memory, 
    \item Relying only on low precision computation (through the Binarized Neural Network concept),
    \item Avoiding multiplication all-together,
    \item The acceptance of some errors without formal error correction.
\end{itemize}

At the same time, \changed{our approach} relies on conventional microelectronics ideas that are  non-biological in nature: 
\begin{itemize}
    \item Relying on fixed point arithmetic to compute sums, whereas brains use analog computation,
    \item  \changed{The use} of sense amplifiers circuits, which are not brain-inspired ,
    \item And the use of a differential structure to reduce errors, a traditional electrical engineering strategy. 
\end{itemize}

Based on these ideas, we designed, fabricated  and tested extensively a memory structure with its \newchanged{peripheral} circuitry, and designed and simulated a full digital system based on \changed{this memory structure}. 
\newchanged{Our results} show  that \newchanged{this} structure allows  implementing neural networks without the use of  Error Correcting Codes, \changed{which} are usually used with emerging memories. \changed{Our approach also} features very attractive properties in terms of energy consumption, and can allow using RRAM devices in ``weak'' programming regime where they have low programming energy and outstanding endurance.  
These results highlight that although in-memory computing cannot \changed{efficiently} rely on Error Correcting Codes, \changed{it can still function without stringent requirements on device variability }
if a differential memory architecture is chosen.

When working on bioinspiration, drawing the line between \newchanged{bio-plausibility} and embracing the differences between \changed{the nanodevices of the brain} and electronic devices is always a challenging question. In this project, we \changed{highlight} that digital electronics can be enriched by \newchanged{biologically}-plausible ideas. When working with nanodevices, it can be beneficial to incorporate device physics questions into the design, and \newchanged{ not to} \changed{necessarily target} the level of determinism that we have been accustomed to by CMOS.

This works opens multiple prospects.
On the  device front,  it could be possible to develop more integrated 2T2R \changed{structures} to increase the density of the memories. The concept \changed{of this work} could also be adapted to other kind of emerging memories, such as phase change memories and spin torque magnetoresistive memories. At the system level, we are now \newchanged{in a}  position to fabricate larger systems, and to investigate \changed{the extension of our concept} to more varied forms of neural network architecture such as convolutional and recurrent ones. 
In the case of convolutional \newchanged{networks}, a dilemma \changed{appears} between \changed{carrying} the in-memory computing approach to its fullest, by replicating physically convolutional kernels, or implementing some sequential computation to minimize resources, as works have started to evaluate already.
These considerations open the way for truly low energy artificial intelligence for both servers and embedded systems.

\section*{Conflict of Interest Statement}
The authors declare that the research was conducted in the absence of any commercial or financial relationships that could be construed as a potential conflict of interest.

\section*{Author Contributions}
EV and EN were in charge of fabrication, and of the initial RRAM characterization. JMP performed the CMOS design of the memory array. MB performed the electrical characterization. TH  designed  the BNN systems. TH, BP and DQ performed the BNN simulations. DQ wrote the initial version of the paper.
JOK, EV, JMP and DQ planned and supervised the project. All authors participated to data analysis, and to the writing of the paper.

\section*{Funding}
This work is supported by the European Research Council Starting Grant NANOINFER (reference: 715872) and the Agence Nationale de la Recherche grant NEURONIC (ANR-18-CE24-0009).



\section*{Data Availability Statement}
The datasets generated for this study are available on request to the corresponding author.

\bibliographystyle{frontiersinSCNS_ENG_HUMS} 
\bibliography{test}

\begin{thebibliography}{49}
\providecommand{\natexlab}[1]{#1}
\expandafter\ifx\csname urlstyle\endcsname\relax
  \providecommand{\doi}[1]{doi:\discretionary{}{}{}#1}\else
  \providecommand{\doi}{doi:\discretionary{}{}{}\begingroup
  \urlstyle{rm}\Url}\fi
\providecommand{\selectlanguage}[1]{\relax}
\providecommand{\bibAnnoteFile}[1]{%
  \IfFileExists{#1}{\begin{quotation}\noindent\textsc{Key:} #1\\
  \textsc{Annotation:}\ \input{#1}\end{quotation}}{}}
\providecommand{\bibAnnote}[2]{%
  \begin{quotation}\noindent\textsc{Key:} #1\\
  \textsc{Annotation:}\ #2\end{quotation}}

\bibitem[{Ambrogio et~al.(2018)Ambrogio, Narayanan, Tsai, Shelby, Boybat, Nolfo
  et~al.}]{ambrogio2018equivalent}
Ambrogio, S., Narayanan, P., Tsai, H., Shelby, R.~M., Boybat, I., Nolfo, C.,
  et~al. (2018).
\newblock Equivalent-accuracy accelerated neural-network training using
  analogue memory.
\newblock \emph{Nature} 558, 60
\bibAnnoteFile{ambrogio2018equivalent}

\bibitem[{Ando et~al.(2017)Ando, Ueyoshi, Orimo, Yonekawa, Sato, Nakahara
  et~al.}]{ando2017brein}
Ando, K., Ueyoshi, K., Orimo, K., Yonekawa, H., Sato, S., Nakahara, H., et~al.
  (2017).
\newblock Brein memory: A 13-layer 4.2 k neuron/0.8 m synapse binary/ternary
  reconfigurable in-memory deep neural network accelerator in 65 nm cmos.
\newblock In \emph{Proc. VLSI Symp. on Circuits} (IEEE), C24--C25
\bibAnnoteFile{ando2017brein}

\bibitem[{Bankman et~al.(2018)Bankman, Yang, Moons, Verhelst, and
  Murmann}]{bankman2018always}
Bankman, D., Yang, L., Moons, B., Verhelst, M., and Murmann, B. (2018).
\newblock An always-on 3.8muj/86 \% cifar-10 mixed-signal binary cnn processor
  with all memory on chip in 28-nm cmos.
\newblock \emph{IEEE Journal of Solid-State Circuits} 54, 158--172
\bibAnnoteFile{bankman2018always}

\bibitem[{Bocquet et~al.(2018)Bocquet, Hirztlin, Klein, Nowak, Vianello, Portal
  et~al.}]{bocquet2018}
Bocquet, M., Hirztlin, T., Klein, J.-O., Nowak, E., Vianello, E., Portal,
  J.-M., et~al. (2018).
\newblock In-memory and error-immune differential rram implementation of
  binarized deep neural networks.
\newblock In \emph{IEDM Tech. Dig.} (IEEE), 20.6.1
\bibAnnoteFile{bocquet2018}

\bibitem[{Chen(2016)}]{chen2016review}
Chen, A. (2016).
\newblock A review of emerging non-volatile memory (nvm) technologies and
  applications.
\newblock \emph{Solid-State Electronics} 125, 25--38
\bibAnnoteFile{chen2016review}

\bibitem[{Chen et~al.(2018)Chen, Li, Lin, Hsu, Li, Yang
  et~al.}]{chen_65nm_2018}
Chen, W.~H., Li, K.~X., Lin, W.~Y., Hsu, K.~H., Li, P.~Y., Yang, C.~H., et~al.
  (2018).
\newblock A 65nm 1mb nonvolatile computing-in-memory {ReRAM} macro with
  sub-16ns multiply-and-accumulate for binary {DNN} {AI} edge processors.
\newblock In \emph{Proc. {ISSCC}}. 494--496.
\newblock \doi{10.1109/ISSCC.2018.8310400}
\bibAnnoteFile{chen_65nm_2018}

\bibitem[{Chen et~al.(2017)Chen, Lin, Lai, Li, Hsu, Lin
  et~al.}]{chen_16mb_2017}
Chen, W.~H., Lin, W.~J., Lai, L.~Y., Li, S., Hsu, C.~H., Lin, H.~T., et~al.
  (2017).
\newblock A 16mb dual-mode {ReRAM} macro with sub-14ns computing-in-memory and
  memory functions enabled by self-write termination scheme.
\newblock In \emph{{IEDM} {Tech}. {Dig}.} 28.2.1--28.2.4.
\newblock \doi{10.1109/IEDM.2017.8268468}
\bibAnnoteFile{chen_16mb_2017}

\bibitem[{Courbariaux et~al.(2016)Courbariaux, Hubara, Soudry, El-Yaniv, and
  Bengio}]{courbariaux2016binarized}
Courbariaux, M., Hubara, I., Soudry, D., El-Yaniv, R., and Bengio, Y. (2016).
\newblock Binarized neural networks: Training deep neural networks with weights
  and activations constrained to+ 1 or-1.
\newblock \emph{arXiv preprint arXiv:1602.02830}
\bibAnnoteFile{courbariaux2016binarized}

\bibitem[{Covi et~al.(2016)Covi, Brivio, Serb, Prodromakis, Fanciulli, and
  Spiga}]{covi2016analog}
Covi, E., Brivio, S., Serb, A., Prodromakis, T., Fanciulli, M., and Spiga, S.
  (2016).
\newblock Analog memristive synapse in spiking networks implementing
  unsupervised learning.
\newblock \emph{Frontiers in neuroscience} 10, 482
\bibAnnoteFile{covi2016analog}

\bibitem[{Dong et~al.(2017)Dong, Kim, Lee, Choi, Li, Wang et~al.}]{dong201711}
Dong, Q., Kim, Y., Lee, I., Choi, M., Li, Z., Wang, J., et~al. (2017).
\newblock 11.2 a 1mb embedded nor flash memory with 39$\mu$w program power for
  mm-scale high-temperature sensor nodes.
\newblock In \emph{2017 IEEE International Solid-State Circuits Conference
  (ISSCC)} (IEEE), 198--199
\bibAnnoteFile{dong201711}

\bibitem[{Editorial(2018)}]{editorial_big_2018}
Editorial (2018).
\newblock Big data needs a hardware revolution.
\newblock \emph{Nature} 554, 145.
\newblock \doi{10.1038/d41586-018-01683-1}
\bibAnnoteFile{editorial_big_2018}

\bibitem[{Faisal et~al.(2008)Faisal, Selen, and Wolpert}]{faisal2008noise}
Faisal, A.~A., Selen, L.~P., and Wolpert, D.~M. (2008).
\newblock Noise in the nervous system.
\newblock \emph{Nature reviews neuroscience} 9, 292
\bibAnnoteFile{faisal2008noise}

\bibitem[{Giacomin et~al.(2019)Giacomin, Greenberg-Toledo, Kvatinsky, and
  Gaillardon}]{giacomin2019robust}
Giacomin, E., Greenberg-Toledo, T., Kvatinsky, S., and Gaillardon, P.-E.
  (2019).
\newblock A robust digital rram-based convolutional block for low-power image
  processing and learning applications.
\newblock \emph{IEEE Transactions on Circuits and Systems I: Regular Papers}
  66, 643--654
\bibAnnoteFile{giacomin2019robust}

\bibitem[{Gregori et~al.(2003)Gregori, Cabrini, Khouri, and
  Torelli}]{gregori2003chip}
Gregori, S., Cabrini, A., Khouri, O., and Torelli, G. (2003).
\newblock On-chip error correcting techniques for new-generation flash
  memories.
\newblock \emph{Proc. IEEE} 91, 602--616
\bibAnnoteFile{gregori2003chip}

\bibitem[{Grossi et~al.(2016)Grossi, Nowak, Zambelli, Pellissier, Bernasconi,
  Cibrario et~al.}]{grossi2016fundamental}
Grossi, A., Nowak, E., Zambelli, C., Pellissier, C., Bernasconi, S., Cibrario,
  G., et~al. (2016).
\newblock Fundamental variability limits of filament-based rram.
\newblock In \emph{IEDM Tech. Dig.} (IEEE), 4--7
\bibAnnoteFile{grossi2016fundamental}

\bibitem[{Grossi et~al.(2018)Grossi, Vianello, Zambelli, Royer, Noel, Giraud
  et~al.}]{grossi2018experimental}
Grossi, A., Vianello, E., Zambelli, C., Royer, P., Noel, J.-P., Giraud, B.,
  et~al. (2018).
\newblock Experimental investigation of 4-kb rram arrays programming conditions
  suitable for tcam.
\newblock \emph{IEEE Transactions on Very Large Scale Integration (VLSI)
  Systems} 26, 2599--2607
\bibAnnoteFile{grossi2018experimental}

\bibitem[{Hsieh et~al.(2017)Hsieh, Chih, Chang, Lin, and
  King}]{hsieh_differential_2017}
Hsieh, W.-T., Chih, Y.-D., Chang, J., Lin, C.-J., and King, Y.-C. (2017).
\newblock Differential {Contact} {RRAM} {Pair} for {Advanced} {CMOS} {Logic}
  {NVM} applications.
\newblock \emph{Proc. SSDM} , 171
\bibAnnoteFile{hsieh_differential_2017}

\bibitem[{Ielmini and Wong(2018)}]{ielmini2018memory}
Ielmini, D. and Wong, H.-S.~P. (2018).
\newblock In-memory computing with resistive switching devices.
\newblock \emph{Nature Electronics} 1, 333
\bibAnnoteFile{ielmini2018memory}

\bibitem[{Indiveri and Liu(2015)}]{indiveri2015memory}
Indiveri, G. and Liu, S.-C. (2015).
\newblock Memory and information processing in neuromorphic systems.
\newblock \emph{Proc. IEEE} 103, 1379--1397
\bibAnnoteFile{indiveri2015memory}

\bibitem[{Ioffe and Szegedy(2015)}]{ioffe2015batch}
Ioffe, S. and Szegedy, C. (2015).
\newblock Batch normalization: Accelerating deep network training by reducing
  internal covariate shift.
\newblock \emph{arXiv preprint arXiv:1502.03167}
\bibAnnoteFile{ioffe2015batch}

\bibitem[{Jouppi et~al.(2017)Jouppi, Young, Patil, Patterson, Agrawal, Bajwa
  et~al.}]{jouppi2017datacenter}
Jouppi, N.~P., Young, C., Patil, N., Patterson, D., Agrawal, G., Bajwa, R.,
  et~al. (2017).
\newblock In-datacenter performance analysis of a tensor processing unit.
\newblock In \emph{Proc. ISCA} (IEEE), 1--12
\bibAnnoteFile{jouppi2017datacenter}

\bibitem[{Kingma and Ba(2014)}]{kingma2014adam}
Kingma, D.~P. and Ba, J. (2014).
\newblock Adam: A method for stochastic optimization.
\newblock \emph{arXiv preprint arXiv:1412.6980}
\bibAnnoteFile{kingma2014adam}

\bibitem[{Klemm and Bornholdt(2005)}]{klemm2005topology}
Klemm, K. and Bornholdt, S. (2005).
\newblock Topology of biological networks and reliability of information
  processing.
\newblock \emph{Proceedings of the National Academy of Sciences} 102,
  18414--18419
\bibAnnoteFile{klemm2005topology}

\bibitem[{Krizhevsky and Hinton(2009)}]{krizhevsky2009learning}
Krizhevsky, A. and Hinton, G. (2009).
\newblock \emph{Learning multiple layers of features from tiny images}.
\newblock Tech. rep., Citeseer
\bibAnnoteFile{krizhevsky2009learning}

\bibitem[{Krizhevsky et~al.(2012)Krizhevsky, Sutskever, and
  Hinton}]{krizhevsky2012imagenet}
Krizhevsky, A., Sutskever, I., and Hinton, G.~E. (2012).
\newblock Imagenet classification with deep convolutional neural networks.
\newblock In \emph{Advances in neural information processing systems}.
  1097--1105
\bibAnnoteFile{krizhevsky2012imagenet}

\bibitem[{Lane et~al.(2016)Lane, Bhattacharya, Georgiev, Forlivesi, Jiao,
  Qendro et~al.}]{lane2016deepx}
Lane, N.~D., Bhattacharya, S., Georgiev, P., Forlivesi, C., Jiao, L., Qendro,
  L., et~al. (2016).
\newblock Deepx: A software accelerator for low-power deep learning inference
  on mobile devices.
\newblock In \emph{Proceedings of the 15th International Conference on
  Information Processing in Sensor Networks} (IEEE Press), 23
\bibAnnoteFile{lane2016deepx}

\bibitem[{LeCun et~al.(1998)LeCun, Bottou, Bengio, and
  Haffner}]{lecun1998gradient}
LeCun, Y., Bottou, L., Bengio, Y., and Haffner, P. (1998).
\newblock Gradient-based learning applied to document recognition.
\newblock \emph{Proc. IEEE} 86, 2278--2324
\bibAnnoteFile{lecun1998gradient}

\bibitem[{Li et~al.(2018)Li, Belkin, Li, Yan, Hu, Ge et~al.}]{li2018efficient}
Li, C., Belkin, D., Li, Y., Yan, P., Hu, M., Ge, N., et~al. (2018).
\newblock Efficient and self-adaptive in-situ learning in multilayer memristor
  neural networks.
\newblock \emph{Nature communications} 9, 2385
\bibAnnoteFile{li2018efficient}

\bibitem[{Lin et~al.(2017)Lin, Zhao, and Pan}]{lin2017towards}
Lin, X., Zhao, C., and Pan, W. (2017).
\newblock Towards accurate binary convolutional neural network.
\newblock In \emph{Advances in Neural Information Processing Systems}. 345--353
\bibAnnoteFile{lin2017towards}

\bibitem[{Ly et~al.(2018)Ly, Grossi, Fenouillet-Beranger, Nowak, Querlioz, and
  Vianello}]{ly2018role}
Ly, D. R.~B., Grossi, A., Fenouillet-Beranger, C., Nowak, E., Querlioz, D., and
  Vianello, E. (2018).
\newblock Role of synaptic variability in resistive memory-based spiking neural
  networks with unsupervised learning.
\newblock \emph{J. Phys. D: Applied Physics}
\bibAnnoteFile{ly2018role}

\bibitem[{Merrikh-Bayat et~al.(2017)Merrikh-Bayat, Guo, Klachko, Prezioso,
  Likharev, and Strukov}]{merrikh2017high}
Merrikh-Bayat, F., Guo, X., Klachko, M., Prezioso, M., Likharev, K.~K., and
  Strukov, D.~B. (2017).
\newblock High-performance mixed-signal neurocomputing with nanoscale
  floating-gate memory cell arrays.
\newblock \emph{IEEE transactions on neural networks and learning systems} 29,
  4782--4790
\bibAnnoteFile{merrikh2017high}

\bibitem[{Natsui et~al.(2018)Natsui, Chiba, and Hanyu}]{natsui2018design}
Natsui, M., Chiba, T., and Hanyu, T. (2018).
\newblock Design of mtj-based nonvolatile logic gates for quantized neural
  networks.
\newblock \emph{Microelectronics journal} 82, 13--21
\bibAnnoteFile{natsui2018design}

\bibitem[{Pedram et~al.(2017)Pedram, Richardson, Horowitz, Galal, and
  Kvatinsky}]{pedram2017dark}
Pedram, A., Richardson, S., Horowitz, M., Galal, S., and Kvatinsky, S. (2017).
\newblock Dark memory and accelerator-rich system optimization in the dark
  silicon era.
\newblock \emph{IEEE Design \& Test} 34, 39--50
\bibAnnoteFile{pedram2017dark}

\bibitem[{Prezioso et~al.(2015)Prezioso, Merrikh-Bayat, Hoskins, Adam,
  Likharev, and Strukov}]{prezioso2015training}
Prezioso, M., Merrikh-Bayat, F., Hoskins, B., Adam, G.~C., Likharev, K.~K., and
  Strukov, D.~B. (2015).
\newblock Training and operation of an integrated neuromorphic network based on
  metal-oxide memristors.
\newblock \emph{Nature} 521, 61
\bibAnnoteFile{prezioso2015training}

\bibitem[{Querlioz et~al.(2015)Querlioz, Bichler, Vincent, and
  Gamrat}]{querlioz2015bioinspired}
Querlioz, D., Bichler, O., Vincent, A.~F., and Gamrat, C. (2015).
\newblock Bioinspired programming of memory devices for implementing an
  inference engine.
\newblock \emph{Proc. IEEE} 103, 1398--1416
\bibAnnoteFile{querlioz2015bioinspired}

\bibitem[{Rastegari et~al.(2016)Rastegari, Ordonez, Redmon, and
  Farhadi}]{rastegari2016xnor}
Rastegari, M., Ordonez, V., Redmon, J., and Farhadi, A. (2016).
\newblock Xnor-net: Imagenet classification using binary convolutional neural
  networks.
\newblock In \emph{Proc. ECCV} (Springer), 525--542
\bibAnnoteFile{rastegari2016xnor}

\bibitem[{Sa{\"\i}ghi et~al.(2015)Sa{\"\i}ghi, Mayr, Serrano-Gotarredona,
  Schmidt, Lecerf, Tomas et~al.}]{saighi2015plasticity}
Sa{\"\i}ghi, S., Mayr, C.~G., Serrano-Gotarredona, T., Schmidt, H., Lecerf, G.,
  Tomas, J., et~al. (2015).
\newblock Plasticity in memristive devices for spiking neural networks.
\newblock \emph{Frontiers in neuroscience} 9, 51
\bibAnnoteFile{saighi2015plasticity}

\bibitem[{Serb et~al.(2016)Serb, Bill, Khiat, Berdan, Legenstein, and
  Prodromakis}]{serb2016unsupervised}
Serb, A., Bill, J., Khiat, A., Berdan, R., Legenstein, R., and Prodromakis, T.
  (2016).
\newblock Unsupervised learning in probabilistic neural networks with
  multi-state metal-oxide memristive synapses.
\newblock \emph{Nature communications} 7, 12611
\bibAnnoteFile{serb2016unsupervised}

\bibitem[{Shafiee et~al.(2016)Shafiee, Nag, Muralimanohar, Balasubramonian,
  Strachan, Hu et~al.}]{shafiee2016isaac}
Shafiee, A., Nag, A., Muralimanohar, N., Balasubramonian, R., Strachan, J.~P.,
  Hu, M., et~al. (2016).
\newblock Isaac: A convolutional neural network accelerator with in-situ analog
  arithmetic in crossbars.
\newblock \emph{ACM SIGARCH Computer Architecture News} 44, 14--26
\bibAnnoteFile{shafiee2016isaac}

\bibitem[{Shih et~al.(2017)Shih, Hsu, Lin, and King}]{shih_twin-bit_2017}
Shih, Y.-H., Hsu, M.-Y., Lin, C.~J., and King, Y.-C. (2017).
\newblock Twin-bit {Via} {RRAM} in 16nm {FinFET} {Logic} {Technologies}.
\newblock \emph{Proc. SSDM} , 137
\bibAnnoteFile{shih_twin-bit_2017}

\bibitem[{Sun et~al.(2018{\natexlab{a}})Sun, Peng, Chen, Liu, Seo, and
  Yu}]{sun2018fully}
Sun, X., Peng, X., Chen, P.-Y., Liu, R., Seo, J.-s., and Yu, S.
  (2018{\natexlab{a}}).
\newblock Fully parallel rram synaptic array for implementing binary neural
  network with (+ 1,- 1) weights and (+ 1, 0) neurons.
\newblock In \emph{Proc. ASP-DAC} (IEEE Press), 574--579
\bibAnnoteFile{sun2018fully}

\bibitem[{Sun et~al.(2018{\natexlab{b}})Sun, Yin, Peng, Liu, Seo, and
  Yu}]{sun2018xnor}
Sun, X., Yin, S., Peng, X., Liu, R., Seo, J.-s., and Yu, S.
  (2018{\natexlab{b}}).
\newblock Xnor-rram: A scalable and parallel resistive synaptic architecture
  for binary neural networks.
\newblock \emph{algorithms} 2, 3
\bibAnnoteFile{sun2018xnor}

\bibitem[{Tang et~al.(2017)Tang, Xia, Li, Wang, and Yang}]{tang2017binary}
Tang, T., Xia, L., Li, B., Wang, Y., and Yang, H. (2017).
\newblock Binary convolutional neural network on rram.
\newblock In \emph{Proc. ASP-DAC} (IEEE), 782--787
\bibAnnoteFile{tang2017binary}

\bibitem[{Wang et~al.(2015)Wang, Ambrogio, Balatti, and Ielmini}]{wang20152}
Wang, Z., Ambrogio, S., Balatti, S., and Ielmini, D. (2015).
\newblock A 2-transistor/1-resistor artificial synapse capable of communication
  and stochastic learning in neuromorphic systems.
\newblock \emph{Frontiers in neuroscience} 8, 438
\bibAnnoteFile{wang20152}

\bibitem[{Wang et~al.(2018)Wang, Joshi, Savel’ev, Song, Midya, Li
  et~al.}]{wang2018fully}
Wang, Z., Joshi, S., Savel’ev, S., Song, W., Midya, R., Li, Y., et~al.
  (2018).
\newblock Fully memristive neural networks for pattern classification with
  unsupervised learning.
\newblock \emph{Nature Electronics} 1, 137
\bibAnnoteFile{wang2018fully}

\bibitem[{Yu(2018)}]{yu2018neuro}
Yu, S. (2018).
\newblock Neuro-inspired computing with emerging nonvolatile memorys.
\newblock \emph{Proc. IEEE} 106, 260--285
\bibAnnoteFile{yu2018neuro}

\bibitem[{Yu et~al.(2016)Yu, Li, Chen, Wu, Gao, Wang et~al.}]{yu2016binary}
Yu, S., Li, Z., Chen, P.-Y., Wu, H., Gao, B., Wang, D., et~al. (2016).
\newblock Binary neural network with 16 mb rram macro chip for classification
  and online training.
\newblock In \emph{IEDM Tech. Dig.} (IEEE), 16--2
\bibAnnoteFile{yu2016binary}

\bibitem[{Zhao et~al.(2009)Zhao, Chappert, Javerliac, and
  Noziere}]{zhao2009high}
Zhao, W., Chappert, C., Javerliac, V., and Noziere, J.-P. (2009).
\newblock High speed, high stability and low power sensing amplifier for
  mtj/cmos hybrid logic circuits.
\newblock \emph{IEEE Transactions on Magnetics} 45, 3784--3787
\bibAnnoteFile{zhao2009high}

\bibitem[{Zhao et~al.(2014)Zhao, Moreau, Deng, Zhang, Portal, Klein
  et~al.}]{zhao2014synchronous}
Zhao, W., Moreau, M., Deng, E., Zhang, Y., Portal, J.-M., Klein, J.-O., et~al.
  (2014).
\newblock Synchronous non-volatile logic gate design based on resistive
  switching memories.
\newblock \emph{IEEE Transactions on Circuits and Systems I: Regular Papers}
  61, 443--454
\bibAnnoteFile{zhao2014synchronous}

\end{thebibliography}





\end{document}